\documentclass[11pt,onecolumn]{article}
\usepackage{a4wide}
\usepackage{mathrsfs}
\usepackage{latexsym,bm}
\usepackage{graphicx}
\usepackage{subfigure}
\usepackage{indentfirst}
\usepackage{slashed}
\usepackage{amsmath}
\usepackage{amssymb}
\usepackage{color}
\usepackage{hyperref}
\usepackage{epsfig}
\usepackage[titletoc]{appendix}
\usepackage{multirow}%
\usepackage{rotating}
\usepackage{cite}
\usepackage{ulem} 
\usepackage[top=1.2cm,bottom=1.5cm,left=2cm,right=2cm]{geometry}
\usepackage{tabularx}
\newcommand{\mL}{\mathcal{L}}

\newcommand{\bra}{\langle}
\newcommand{\ket}{\rangle}
\newcommand{\nn}{\nonumber}

\newcommand{\mT}{\mathcal{T}}
\newcommand{\mK}{\mathcal{K}}
\newcommand{\uao}{U_A(1)}

\begin{document}
\thispagestyle{empty}
\title{ \Large \bf Thermal behaviors of light scalar resonances at low temperatures }
\author{\small Rui Gao$^a$,~ Zhi-Hui Guo$^a$\thanks{Corresponding author: zhguo@hebtu.edu.cn},~ Jin-Yi Pang$^{b,c}$\thanks{Corresponding author: pang@hiskp.uni-bonn.de}  \\[0.3em] 
{ \small\it ${}^a$  Department of Physics and Hebei Advanced Thin Films Laboratory, } \\
{\small\it Hebei Normal University,  Shijiazhuang 050024, China}\\ 
{\small\it ${}^b$ College of Science, University of Shanghai for Science and Technology, } \\
{\small\it Jungong Rd. 334, 200093 Shanghai, China }\\
{\small\it  ${}^c$ Helmholtz-Institut f\"ur Strahlen- und Kernphysik and Bethe Center }\\
{\small\it for Theoretical Physics, Universit\"at Bonn, D--53115 Bonn, Germany}
}
\date{}

\maketitle

\begin{abstract} 
We study the thermal properties of the lowest multiplet of the QCD light-flavor scalar resonances, including the  $f_0(500)/\sigma$, $K_{0}^{*}(700)/\kappa$, $f_0(980)$ and $a_0(980)$, in the framework of unitarized $U(3)$ chiral perturbation theory. After the successful fits to the meson-meson scattering inputs, such as the phase shifts and inelasticities, we obtain the unknown parameters and further calculate the resonance poles and their residues at zero temperature. By including the finite-temperature effects in the unitarized meson-meson scattering amplitudes, the thermal behaviors of the scalar resonance poles in the complex energy plane are studied. The masses of $\sigma$ and $\kappa$ are found to considerably decrease when increasing the temperatures, while their widths turn out to be still large when the temperatures reach around $200$~MeV. In contrast, both the masses and widths of the $f_0(980)$ and $a_0(980)$ are only slightly changed. 
\end{abstract}

\section{Introduction}
Identifying the pattern of the chiral symmetry restoration, which plays the key role in understanding the complex phenomena from the relativistic heavy ion collisions, is one of the most important subjects in the study of QCD phase diagram. The restoration of the chiral symmetry will definitely modify the hadronic spectrum at finite temperatures, which in turn will affect the hadron yields measured in the heavy-ion-collision experiments. E.g., it is found that the inclusion of the broad scalar resonance $f_0(500)$ (also named as  $\sigma$) in the hadron-resonance-gas model clearly improves the description of the experimental data~\cite{Andronic:2008gu}. 

In this work we focus on the thermal behaviors of the lowest multiplet of the light-flavor QCD scalar resonances, including the $\sigma$, $f_0(980)$, $K_0^*(700)$ (also named as $\kappa$) and $a_0(980)$. As the lightest QCD scalar resonance and sharing the same quantum numbers as the vacuum, $\sigma$ has been extensively studied both at zero and finite temperatures~\cite{Pelaez:2015qba}. After decades of precise and rigorous dispersive studies, it is now recognized in PDG that the uncertainties of the mass and width of the broad $\sigma$ resonance reach the precisions of several tens of MeV. For such a broad resonance, it is not appropriate to still use the conventional Breit-Wigner formalism both in the vacuum and at finite temperatures. Instead the inverse-amplitude-method (IAM) up to the one-loop level has been employed to investigate the thermal properties of the $\sigma$ in a series of papers in Refs.~\cite{Dobado:2002xf,FernandezFraile:2007fv,Cabrera:2008tja,Cortes:2015emo}. It is also found that around the transition temperature $T_c$ the inclusion of the thermal $\sigma$ poles in the scalar susceptibilities can develop a maximum, which is consistent with the results in the lattice study~\cite{Nicola:2013vma,GomezNicola:2017bhm}. 

Instead of including further the higher order corrections in the chiral amplitudes, we proceed the discussions by simultaneously studying all the members of the possible lowest multiplet of the light scalar resonances $\sigma$, $f_0(980)$, $\kappa$ and $a_0(980)$ within the unitarized chiral perturbation theory ($\chi$PT). Through this exploratory study, we obtain the thermal behaviors of all the aforementioned resonance poles, which can provide useful guides for the hadron-resonance-gas models and gain insights of the mechanism of the chiral symmetry restoration.

The article is organized as follows. Sec.~\ref{sec.lagrangian} is devoted to the discussions of the relevant $S$-wave chiral amplitudes and their fits to the scattering inputs. The resulting resonance poles and residues at zero temperature are also given in this section. The thermal trajectories of the scalar resonance poles at finite temperatures will be then discussed in detail in Sec.~\ref{sec.ft}. Finally we give a short summary and conclusions in Sec.~\ref{sec.conclusion}.

\section{Unitarized S-wave chiral amplitudes and scalar resonances at zero temperature}\label{sec.lagrangian}

Meson-meson scattering provides an important approach to study the resonance dynamics, where the hadron resonances correspond to the poles in the complex energy plane of the scattering amplitudes. E.g. $\sigma$ and $f_0(980)$ appear in the $\pi\pi$ and $K\bar{K}$ coupled-channel scattering with $(I,J)=(0,0)$, being $I$ the isospin quantum number and $J$ the angular momentum. The most relevant channel for $\kappa$ is the $K\pi$ scattering with $(I,J)=(1/2,0)$, and $a_0(980)$ naturally appears in the $\pi\eta$ and $K\bar{K}$ scattering with $(I,J)=(1,0)$. Since the $\chi$PT relies on the perturbative expansions of the external momenta and light-flavor quark masses~\cite{Weinberg:1978kz,Gasser:1983yg,Gasser:1984gg}, it is impossible to generate resonances from the  perturbative $\chi$PT scattering amplitudes alone. It is evident that the combination of the $\chi$PT and unitarity offers an efficient way to study the aforementioned scalar resonances~\cite{Dobado:1989qm,Oller:1997ti,Dobado:1996ps,Oller:1998hw,Oller:1998zr,Dai:2011bs}. 

The up-to-date perturbative meson-meson scattering amplitudes at zero temperature have been calculated up to two loops for the three-flavor $\chi$PT~\cite{Bijnens:2004bu,Bijnens:2004eu}. At finite temperatures, the perturbative meson-meson amplitudes have only been calculated up to the one-loop level for the two-flavor $\chi$PT~\cite{GomezNicola:2002tn}. The one-loop calculation of the meson-meson scattering amplitudes in the three-flavor $\chi$PT at finite temperatures is still missing and clearly deserves an independent work. According the previous works~\cite{Oller:1997ti,Oller:1998hw,Oller:1998zr,Guo:2016zep}, both the relevant experimental data and the lattice energy levels of the meson-meson scattering below and around 1~GeV in the scalar channels can be well reproduced by taking the leading order (LO) $\chi$PT amplitudes in the unitarization approach. The resulting masses and widths of the scalar resonances from such studies look quite reasonable and are quantitatively compatible with the various rigorous dispersive results~\cite{PDG}. It is plausible that the main features of thermal properties of scalar resonances can be also obtained in such a approach. Therefore, in the following discussions, we will take the leading order perturbative $\chi$PT amplitudes and includes the finite-temperature effects through the unitarization procedure. 

We follow Refs.~\cite{Guo:2011pa,Guo:2012ym,Guo:2012yt} to include the perturbative LO meson-meson scattering from $U(3)$ $\chi$PT. To set up the notations, we simply recapitulate the main results below. The LO $U(3)$ $\chi$PT Lagrangian includes three terms 
\begin{eqnarray}\label{eq.laglo}
\mL= \frac{ F^2}{4}\bra u_\mu u^\mu \ket+
\frac{F^2}{4}\bra \chi_+ \ket
+ \frac{F^2}{3}M_0^2 \ln^2 \det u \,,
\end{eqnarray}
where the chiral building blocks are given by
\begin{eqnarray}\label{defbb}
&& U =  u^2 = e^{i\frac{ \sqrt2\Phi}{ F}}\,, \qquad \chi = 2 B (s + i p) \,,\qquad \chi_\pm  = u^\dagger  \chi u^\dagger  \pm  u \chi^\dagger  u \,, \nn\\
&& u_\mu = i u^\dagger  D_\mu U u^\dagger \,, \qquad  D_\mu U \, =\, \partial_\mu U - i (v_\mu + a_\mu) U\, + i U  (v_\mu - a_\mu) \,,
\end{eqnarray}
and the $U(3)$ matrix of the pseudo Nambu-Goldstone bosons (pNGBs) reads
\begin{equation}\label{phi1}
\Phi \,=\, \left( \begin{array}{ccc}
\frac{1}{\sqrt{2}} \pi^0+\frac{1}{\sqrt{6}}\eta_8+\frac{1}{\sqrt{3}} \eta_0 & \pi^+ & K^+ \\ \pi^- &
\frac{-1}{\sqrt{2}} \pi^0+\frac{1}{\sqrt{6}}\eta_8+\frac{1}{\sqrt{3}} \eta_0   & K^0 \\  K^- & \overline{K}^0 &
\frac{-2}{\sqrt{6}}\eta_8+\frac{1}{\sqrt{3}} \eta_0
\end{array} \right)\,.
\end{equation}
$F$ is the LO pion decay constant, with the normalization $F_\pi=92.1$~MeV. The last term in Eq.~\eqref{eq.laglo} includes the contribution from the QCD $\uao$ anomaly, which gives the singlet $\eta_0$ the LO mass $M_0$.

For the sake of completeness, in the Appendix A we provide the explicit formulas of the LO $S$-wave $U(3)$ meson-meson scattering amplitudes $T_{IJ}(s)$, which were calculated in Ref.~\cite{Guo:2011pa}. The LO amplitudes given by Eq.~\eqref{eq.laglo} only include the contact interactions, which do not contain any crossed-channel cut. The on-shell partial-wave scattering amplitude in the elastic case can be written as~\cite{Oller:1998zr} 
\begin{eqnarray} \label{eq.ut}
\mT_{IJ}(s) = \dfrac{\mK(s)}{1 - \mK(s)\,G(s)}\,, 
\end{eqnarray}
where $\mK(s)$ will be given by the LO $S$-wave $U(3)$ $\chi$PT amplitudes $T_{IJ}(s)$ in this work and the function $G(s)$ includes nonperturbatively the contribution from the right-hand cut. When the higher order $\chi$PT contributions, including the chiral loops, are included, the formalism in Eq.~\eqref{eq.ut} is still valid and can be matched to the perturbative $\chi$PT at low energies to obtain the proper $\mK(s)$ function~\cite{Oller:1999me,Oller:2000fj}. The essential idea is that by construction the function $\mK(s)$ only contains the crossed-channel contributions,  including the contact terms, and the function $G(s)$ only includes the right-hand cut. In such a way, the prescription of Eq.~\eqref{eq.ut} can be regarded as an algebraic approximation of the N/D method~\cite{Oller:1999me,Oller:2000fj}. Explicit examples of the unitarization of the one-loop $U(3)$ $\chi$PT have been given in Refs.~\cite{Guo:2011pa,Guo:2012ym,Guo:2012yt}. 

The two-body unitarity requires that 
\begin{equation}\label{eq.img}
{\rm Im}G(s) = \rho(s) \,\theta(s-s_{\rm th})  \equiv  \dfrac{q(s)}{8\pi\sqrt{s}}\,\theta(s-s_{\rm th}) \,,
\end{equation}
where $s_{\rm th}$ denotes the threshold, $\theta(x)$ is the Heaviside step function and the three momenta in the center of mass (CM) frame is given by 
\begin{equation}\label{eq.q3}
q(s) = \dfrac{\sqrt{[s-(m_1+m_2)^2][s-(m_1-m_2)^2]}}{2\sqrt{s}}\,,
\end{equation}
with $m_1$ and $m_2$ the masses of the two particles in question. 
Next one can use the imaginary part of the function $G(s)$ to build a once subtracted dispersion relation to get the analytical expression of $G(s)$. Alternatively, one can also use the dimensional regularization to calculate  the $G(s)$ function via 
\begin{eqnarray}\label{eq.defg}
G(s)=-i\int\frac{{\rm d}^4k}{(2\pi)^4}
\frac{1}{(k^2-m_{1}^2+i\epsilon)[(P-k)^2-m_{2}^2+i\epsilon ]}\ ,\qquad
s\equiv P^2\ \,,
\end{eqnarray}
which explicit expression takes the form by replacing the divergent term with a constant~\cite{Oller:1998zr}
\begin{eqnarray}\label{eq.gfunc}
G(s)^{\rm DR} &=& -\frac{1}{16\pi^2}\left[ a(\mu^2) + \log\frac{m_2^2}{\mu^2}-x_+\log\frac{x_+-1}{x_+}
-x_-\log\frac{x_--1}{x_-} \right]\,, 
\end{eqnarray}
where $\mu$ denotes the regularization scale and $x_\pm$ are defined as 
\begin{equation}
 x_\pm =\frac{s+m_1^2-m_2^2}{2s}\pm \frac{q(s)}{\sqrt{s}}\,.
\end{equation}
One should notice that the function $G(s)$ is independent of the scale $\mu$, due to the cancellation of the $\mu$ dependences of the first and second terms in Eq.~\eqref{eq.gfunc}. In the following discussion we will fix $\mu=770$~MeV throughout. Notice that there is a minus sign difference between the $G(s)$ function in Eq.~\eqref{eq.gfunc} and the one in Refs.~\cite{Guo:2011pa,Guo:2012yt}, which is compensated by the minus sign in the denominator of the unitarized amplitude~\eqref{eq.ut}, so that the imaginary part of the $G(s)$ is positive. 

For the coupled-channel scattering, the entries of $\mK(s)$ and $G(s)$ in Eq.~\eqref{eq.ut} should be understood as matrices spanned in the channel space. For the case with definite isospin and angular momentum, $G(s)$ corresponds to a diagonal matrix and its diagonal elements can be calculated via Eq.~\eqref{eq.gfunc} by using the proper masses in question. There are five coupled channels in the $(I,J)=(0,0)$ case, including  $\pi\pi$, $K\bar{K}$, $\eta\eta$, $\eta\eta'$ and $\eta'\eta'$. Three relevant channels enter in the $(I,J)=(1/2,0)$ and $(I,J)=(1,0)$ cases, which are the $K\pi, K\eta, K\eta'$ and $\pi\eta$, $K\bar{K}$ and $\pi\eta'$, respectively. In the $U(3)$ $\chi$PT, the massive $\eta'$ state is explicitly included, which however plays a marginal role in the study of the low lying scalar resonances $\sigma$, $f_0(980)$, $\kappa$ and $a_0(980)$~\cite{Guo:2011pa,Guo:2012ym,Guo:2012yt}. In contrast, for the excited scalar resonances with higher masses, it is evident that their couplings to the $\eta'$ state become large~\cite{Guo:2011pa,Guo:2012ym,Guo:2012yt}. 

The $S$ matrix is related to the unitarized $\mT$ amplitude in Eq.~\eqref{eq.ut} via 
\begin{equation}
 S  = 1 + 2 i \sqrt{\rho(s)}\cdot \mT(s)\cdot \sqrt{\rho(s)}\,.  
\end{equation} 
In the coupled-channel case, $\rho(s)$ should be understood as diagonal matrix and its non-vanishing elements can be calculated through Eqs.~\eqref{eq.img} and \eqref{eq.q3}. The phase shifts $\delta_{kk}, \delta_{kl}$ and the inelasticities  $\varepsilon_{kk}, \varepsilon_{k l}$, with $k\neq l$, can be obtained with the matrix elements $S_{kk}$ and $S_{k l}$ 
\begin{align}\label{eq.defsmat} 
S_{ k k} = \varepsilon_{k k} {\rm e}^{2 i \delta_{ k k}}\,, \qquad 
S_{ k l} = i \varepsilon_{k l} {\rm e}^{ i \delta_{ k l}}\,.
\end{align}
The inelasticities $\varepsilon_{kk}$ fulfill the condition $0\leq \varepsilon_{kk}\leq 1$. 

We use the physical masses for the $\pi,K,\eta,\eta'$ and the physical value $F_\pi$ in the LO scattering amplitudes. According to the Lagrangian in Eq.~\eqref{eq.laglo}, the LO $\eta$-$\eta'$ mixing angle $\theta$ is given by~\cite{Guo:2011pa}
\begin{eqnarray}\label{eq.loangle}
\sin{\theta} = -\left( \sqrt{1 +
\frac{ \big(3M_0^2 - 2\Delta^2 +\sqrt{9M_0^4-12 M_0^2 \Delta^2 +36 \Delta^4 } \big)^2}{32 \Delta^4} } ~\right )^{-1}\,, 
\end{eqnarray}
where $\Delta^2 = \overline{m}_K^2 - \overline{m}_\pi^2$, and $\overline{m}_K$ and $\overline{m}_\pi$ are
the LO kaon and pion masses, in order. We will estimate $\overline{m}_K$ and $\overline{m}_\pi$ by their corresponding physical values. For the LO mass $M_0$ of the $\eta_0$, we will take the value $M_0=820$~MeV that has been recently determined in Ref.~\cite{Gu:2018swy} by fitting the updated lattice data of the $\eta$-$\eta'$ mixing.

\begin{table}[htbp]
\centering
\begin{tabular}{ c c c c }
\hline\hline
$\pi\pi$ with $(I,J)=(0,0)$ & $a_{SL,1}$     &$a_{SL,2}$       & $\chi^2/d.o.f$ 
\\ 
               & $-1.13^{+0.19}_{-0.17}$ & $-1.93^{+0.23}_{-0.29}$ &  149.0/(95-2)
\\ \hline
$K\pi$ with  $(I,J)=(1/2,0)$ & $a_{SL,1}$    &    $\chi^2/d.o.f$  &
\\ 
               & $-0.42^{+0.16}_{-0.16}$ &   16.2/(36-1)  &               
\\ 
\hline\hline
\end{tabular}
\caption{The values of the subtraction constants from the fits. In the $\pi\pi$ scattering with $(I,J)=(0,0)$, $a_{SL,1}$ and $a_{SL,2}$ correspond to the subtraction constants in the $\pi\pi$ and $K\bar{K}$ channels, respectively. For the remaining channels $\eta\eta, \eta\eta'$ and $\eta'\eta'$, we fix their subtraction constants as the same as $a_{SL,1}$. For other possibilities to perform the fits, see the text for details. In the $K\pi$ scattering with $(I,J)=(1/2,0)$, we take the same value of the subtraction constant for all the three coupled channels. For the $\pi\eta, K\bar{K}$ and $\pi\eta'$ coupled-channel  scattering, we take the universal subtraction constant $a_{SL,1}=-1.44\pm 0.15$ for all the three channels as determined in Ref.~\cite{Guo:2016zep}. }\label{tab.fit}
\end{table}

\begin{figure}[htbp]
   \centering
   \begin{minipage}[t]{0.48\textwidth}
      \centering
   \includegraphics[width=0.99\textwidth,angle=-0]{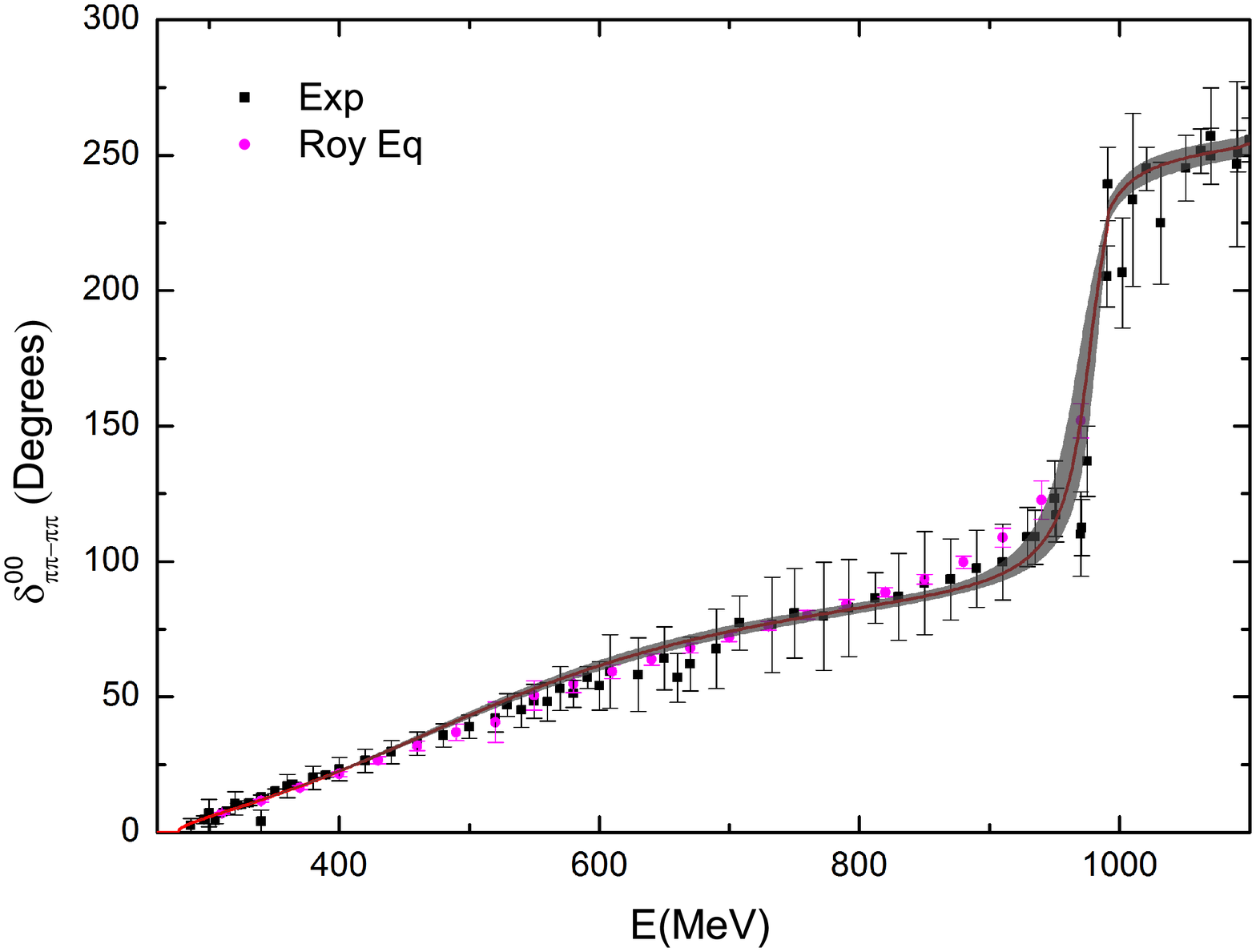} 
   \end{minipage}
      \begin{minipage}[t]{0.49\textwidth}
         \centering
   \includegraphics[width=0.99\textwidth,angle=-0]{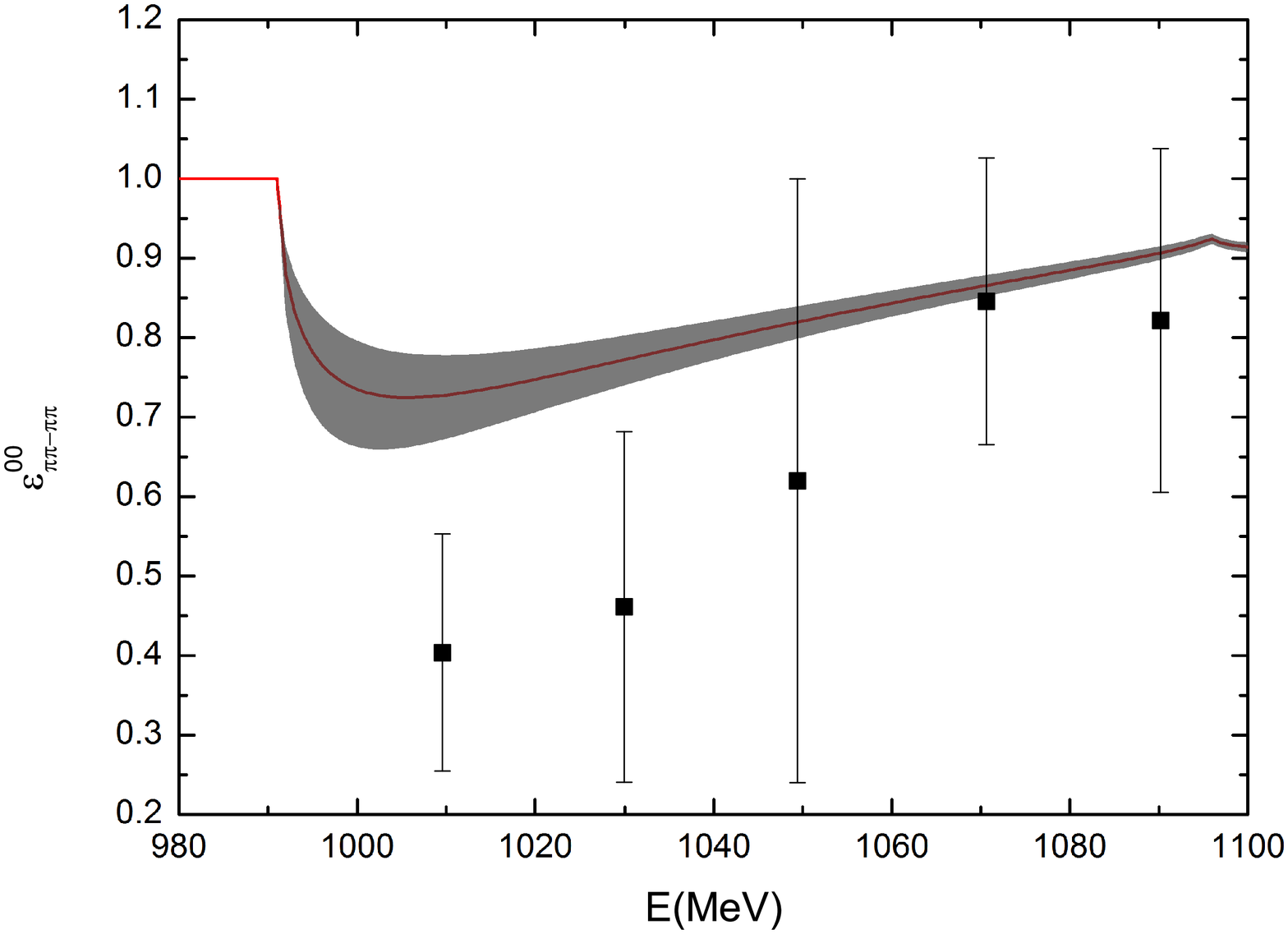} 
   \end{minipage}
  \caption{ Reproduction of the phase shifts (left panel) and inelasticities (right panel) of the $\pi\pi$ scattering with $(I,J)=(0,0)$. The experimental data correspond to those used in Refs.~\cite{Oller:1998hw,Guo:2011pa,Guo:2012ym,Guo:2012yt}, which average various data points in Ref.~\cite{pipidata}. The precise data from the Roy equation analysis are taken from Ref.~\cite{GarciaMartin:2011cn}. The shaded areas denote our estimates of the theoretical uncertainties at the one-sigma level.   }
   \label{fig.fit00}
\end{figure}

\begin{figure}[htbp]
   \centering
   \includegraphics[width=0.5\textwidth,angle=-0]{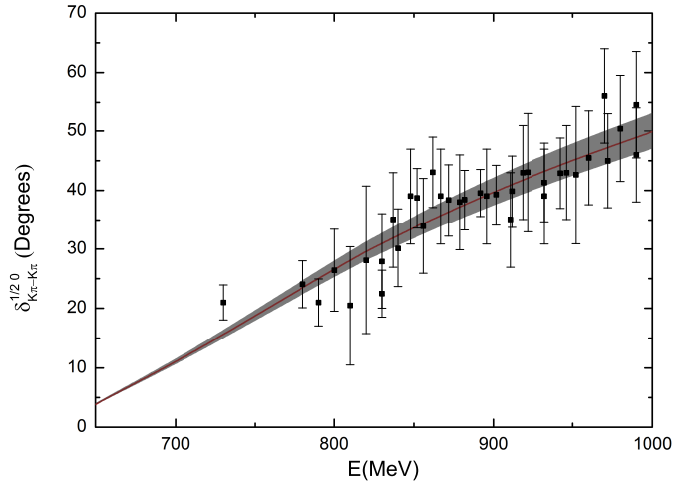} 
  \caption{  Reproduction of the phase shifts of the $K\pi$ scattering with $(I,J)=(1/2,0)$. The experimental data are taken from Ref.~\cite{pikdata}. The shaded area corresponds to the theoretical uncertainties at the one-sigma level.  }
   \label{fig.fit1d20}
\end{figure}

The remaining unknown parameters in the unitarized scattering amplitudes $\mT(s)$ in Eq.~\eqref{eq.ut} are the subtraction constants, which will be determined in the fits to the phase shifts and inelasticities for the $\pi\pi$ scattering with $(I,J)=(0,0)$ and the $K\pi$ scattering with $(I,J)=(1/2,0)$. Since only the LO perturbative amplitudes are included, we include the experimental data for the $\pi\pi$ up to 1100~MeV and the $K\pi$ up to 1000~MeV in the fits. In addition to the experimental data used in Refs.~\cite{Oller:1998hw,Guo:2011pa,Guo:2012ym,Guo:2012yt}, we also take into account the precise isoscalar and scalar $\pi\pi$ phase shifts determined from the Roy equation~\cite{GarciaMartin:2011cn}. The reproductions of the data for the $\pi\pi$ scattering with $(I,J)=(0,0)$ and the $K\pi$ scattering with $(I,J)=(1/2,0)$ are given in Figs.~\ref{fig.fit00} and \ref{fig.fit1d20}, respectively. The resulting values of the subtraction constants are summarized in Table~\ref{tab.fit}. It is remarkable that with one and two free parameters in the $K\pi$ and $\pi\pi$ scattering cases, respectively, one can well reproduce the relevant data from the experiments and Roy equation. We have also tried other ways to perform the fits for the $\pi\pi$ with $(I,J)=(0,0)$. E.g., to fix the subtraction constants of the $\eta\eta, \eta\eta'$ and $\eta'\eta'$ channels as the one from the $K\bar{K}$ channel, instead of the $\pi\pi$ case in Table~\ref{tab.fit}, the fits will get slightly worse. Freeing the subtraction constants in the $\eta\eta, \eta\eta'$ and $\eta'\eta'$ channels will improve the fits, but the resulting values of the subtraction constants, which bear large uncertainties, do not seem to fall in the reasonable ranges. In all the three cases, it turns out that the resonances in the scattering amplitudes are more or less compatible and we will focus on the fits shown in Table~\ref{tab.fit} in later discussions. For the $\pi\eta, K\bar{K}$ and $\pi\eta'$ coupled-channel scattering, the direct experimental measurements on the scattering processes are still absent, instead the amplitudes are determined by  fitting the lattice finite-volume energy levels in Ref.~\cite{Guo:2016zep}. We will take the subtraction constants determined in the former reference in this work.

After the determination of all the unknown parameters, we now discuss the resonances, corresponding to the poles in the complex energy plane, in the unitarized scattering amplitudes. 
The analytical continuation of the unitarized amplitudes in Eq.~\eqref{eq.ut} can be easily done by properly extrapolating the $G(s)$ function in Eq.~\eqref{eq.gfunc} to the complex energy plane. Two Riemann sheets (RS's) for the $G(s)$ function can be defined. On the unphysical/second RS it reads~\cite{Oller:1997ti} 
\begin{eqnarray}\label{eq.defg2ndrs} 
G(s)^{\rm DR}_{\rm II}(s) = G(s)^{\rm DR} - i \frac{q(s)}{4\pi \sqrt{s}}\,,
\end{eqnarray}
with the $G(s)^{\rm DR}$ on the physical/first RS given in Eq.~\eqref{eq.gfunc}. By combing Eqs.~\eqref{eq.img} and  \eqref{eq.defg2ndrs}, it is clear that along the real $s$ axis above the threshold the imaginary parts of the $G(s)$ function on the first and second RS's have opposite signs. As a result, $2^n$ RS's can be defined for the $n$-channel scattering problem. We  denote the first, second, third and fourth RS's as $(+,+,+,\cdots,+)$, $(-,+,+,\cdots,+)$, $(-,-,+,\cdots,+)$ and $(+,-,+,\cdots,+)$, respectively. The entries of plus and minus symbols correspond to the signs of the imaginary parts of the $G(s)$ functions in different channels. The residues $\gamma$ at the resonance pole $s_P$, which denote the coupling strengths of the resonance, are given by   
\begin{eqnarray}
\mT(s)= -\lim_{s\to s_P}\frac{\gamma\gamma^{\rm T}}{s-s_P}\,,
\end{eqnarray}
where $\gamma$ is an $n$-row vector and its transpose is $\gamma^{\rm T}=(\gamma_1,\gamma_2,\cdots,\gamma_n)$. 
The resonance poles and their residues at zero temperature are collected in Table~\ref{tab.polezerot}. The successful reproduction of the input data and the quantitative agreements of the resonance poles in Table~\ref{tab.polezerot} with those estimated in PDG~\cite{PDG} provide us a confident starting point to extend the current discussions of the light scalar resonances to the finite temperatures. 

\begin{table}[htbp]
\centering
\begin{scriptsize}
\begin{tabular}{ c c c c c c c}
\hline\hline
 $R$ & M(MeV)   & Width/2(MeV) & $\left|\gamma_1\right|$(GeV) &Ratios
\\ \hline
$\sigma$ & $465^{+1}_{-2}$ & $234^{+ 8}_{-8}$ & $3.14^{+0.03}_{-0.03}$&$0.45^{+0.01}_{-0.01}(K\bar{K}/\pi\pi)$&$0.02^{+0.02}_{-0.01}(\eta\eta/\pi\pi)$\\&&&&$0.067^{+0.007}_{-0.007}(\eta\eta'/\pi\pi)$&
$0.06^{+0.01}_{-0.02}(\eta'\eta'/\pi\pi)$
\\ \hline
$f_{0}(980)$ & $977^{+6}_{-9}$ &  $15^{+5}_{-3}$ &  $1.29^{+0.19}_{-0.15}$&$3.05^{+0.64}_{-0.57}(K\bar{K}/\pi\pi)$&$2.23^{+0.56}_{-0.47}(\eta\eta/\pi\pi)$\\&&&&
$1.06^{+0.20}_{-0.19}(\eta\eta'/\pi\pi)$&$1.10^{+0.24}_{-0.21}(\eta'\eta'/\pi\pi)$
\\ \hline
$\kappa$ & $738^{+ 8}_{-9}$ & $274^{+ 8}_{-9}$ & $4.22^{+0.06}_{-0.07}$&  $0.46^{+0.02}_{-0.02}(K\eta/K\pi)$&$0.39^{+0.01}_{-0.02}(K\eta'/K\pi)$
\\ \hline
$a_{0}(980)$&$1037^{+17}_{-14}$&$44^{+6}_{-9}$&$3.8^{+0.3}_{-0.2}$&$1.43^{+0.03}_{-0.03}(K\bar{K}/\pi\eta)$&$0.05^{+0.01}_{-0.01}(\pi\eta'/\pi\eta)$
\\
\hline\hline
\end{tabular}
\caption{The masses, widths and residues of various resonances at zero temperature in the Second RS. $\gamma_1$ denotes the residue of the lightest channel of each resonance. The values in the last two columns correspond to the ratios $|\gamma_i/\gamma_1|$.  }\label{tab.polezerot}
\end{scriptsize}
\end{table}

\section{The scalar resonances at finite temperatures}~\label{sec.ft}

In the framework of $\chi$PT, the chiral loops will introduce the finite-temperature effects, while the tree-level Feynman diagrams are free of the finite-temperature corrections~\cite{Gasser:1986vb,Gerber:1988tt,GomezNicola:2002tn}. This implies that in the present work the LO partial-wave scattering amplitudes will not get modified when including the finite temperatures. It is the $G(s)$ function incorporated through the unitarization procedure that will introduce the finite-temperature corrections. A similar theoretical approach has been recently applied to the study of charmed mesons in Refs.~\cite{Cleven:2017fun,Cleven:2019cre}. We point out that in addition to the $s$-channel unitarity loops, the thermal corrections from the crossed-channel and chiral tadpole loops could be also relevant in the study of the resonance properties at finite temperatures. Although the tadpole loop diagrams share the same forms both at zero and finite temperatures, the complete finite-temperature calculation of the full contributions in three-flavor $\chi$PT clearly deserves another independent work.  At zero temperature our study in the previous section and many other works, such as Refs.~\cite{Oller:1997ti,Oller:1998hw,Oller:1998zr,Guo:2016zep}, has shown that the scalar resonances $\sigma$, $\kappa$, $a_0(980)$ and $f_0(980)$ can be well described by only unitarizing the LO contact amplitudes from $\chi$PT, indicating that the $s$-channel unitarity plays the dominant role in the scalar resonance dynamics. The exploratory study in this work assumes that maybe it also holds for the study at finite temperatures, or at least the unitarzation of the LO $\chi$PT may give a qualitatively correct description of the thermal trajectories of the scalar resonances below 1~GeV. 

In this work, we use the imaginary time formalism to include the finite-temperature corrections~\cite{LeBellac}. 
Although it is a standard problem to calculate the loop function of Eq.~\eqref{eq.defg} at finite temperatures~\cite{LeBellac}, we give a practical derivation of the explicit formula in the Appendix B. In the CM frame with real energy squared $s$, the expression of the finite-temperature corrections to the $G(s)$ function in Eq.~\eqref{eq.defg} for $T\neq 0$ takes the form 
\begin{eqnarray} \label{eq.gft}
G(s)^{T\neq 0}&=& \int_{0}^{\infty} \dfrac{k^2 dk}{8\pi^2 E_1 E_2} \bigg\{ \dfrac{1}{E+E_1+E_2}\bigg[ f(E_1)+ f(E_2) \bigg] + \dfrac{1}{E+E_1-E_2}\bigg[ -f(E_1)+f(E_2)\bigg]
\nonumber \\ &&
+ \dfrac{1}{E-E_1+E_2}\bigg[ f(E_1)-f(E_2)\bigg] \bigg\} 
- {\rm P.V.}\int_{0}^{\infty} \dfrac{k^2 dk}{8\pi^2 E_1 E_2}  \dfrac{1}{E-E_1-E_2}\bigg[ f(E_1)+f(E_2)\bigg]
\nonumber \\ &&
+ \dfrac{i q(s)}{8\pi E} \bigg[ f(\widetilde{E}_1)  + f(\widetilde{E}_2) \bigg]\,\theta(s-s_{\rm th})\,,
\end{eqnarray}
with $s=E^2, \,\beta=1/T, \,E_i=\sqrt{k^2+m_i^2}, \,\widetilde{E}_i=\sqrt{q(s)^2+m_i^2}$, $q(s)$ the magnitude of the on-shell three momenta in the CM frame and the standard Bose distribution function $f(x)$ given by
\begin{eqnarray}
f(x)=  \dfrac{1}{e^{\beta x}-1}\,.
\end{eqnarray}
For the first three integrals of Eq.~\eqref{eq.gft}, they are regular in the physical region and can be easily calculated numerically. The fourth term with the symbol P.V. corresponds to taking the principal value of the integral. The last term of Eq.~\eqref{eq.gft} denotes the imaginary part of the thermal corrections to the $G(s)$ function in the energy region above threshold. For the details of the calculation of the expression of Eq.~\eqref{eq.gft}, we refer to the Appendix B. It is stressed that below the threshold there are the so-called thermal Landau cuts, as pointed out in Refs.~\cite{Weldon:1983jn,Ghosh:2009bt,Nicola:2014eda}. In the CM frame of the thermal two-body scattering, the Landau cuts can extend to the positive real $s$ axis up to $(m_1-m_2)^2$. The expressions of the imaginary parts along the Landau cuts are not explicitly shown in Eq.~\eqref{eq.gft}. Within the on-shell approximation of the unitarization approach in Eq.~\eqref{eq.ut}, we do not expect that the Landau cuts will have important influences, as long as the thermal resonance poles are not close to those cuts, which are indeed the common cases in our study, as discussed in detail later.

The unitarized amplitude at finite temperature $T$ reads 
\begin{eqnarray}\label{eq.utft}
\mT^{\rm FT}(s) = \bigg[1 - \mK(s)\cdot G(s)^{\rm FT} \bigg]^{-1}\cdot \mK(s)\,,
\end{eqnarray}
where the corrected $G(s)$ with the finite-temperature effect is 
\begin{eqnarray}\label{eq.gftsum}
 G(s)^{\rm FT} = G(s)^{\rm DR} + G(s)^{T\neq 0} \,,
\end{eqnarray}
with $G(s)^{\rm DR}$ and $G(s)^{T\neq 0}$ given in Eqs.~\eqref{eq.gfunc} and \eqref{eq.gft}, respectively. Similar as the zero-temperature case, $\mK(s)$ and $G(s)^{\rm FT}$ should be understood as matrices in the coupled-channel scattering. Comparing with the zero-temperature amplitudes in Eq.~\eqref{eq.ut}, no additional free parameters are introduced to the amplitudes at $T\neq 0$ in Eq.~\eqref{eq.utft}. Therefore the thermal behaviors of the unitarized amplitudes and the scalar resonances will be pure predictions in the unitarized $\chi$PT. 
In order to study the thermal trajectories of the resonance poles, we need to first perform the analytical continuation of the unitarized amplitude $\mT^{\rm FT}(s)$ to the unphysical RS and then search the poles in the complex energy plane. By taking into account that the signs of the imaginary parts of $G(s)^{\rm FT}$ on the first and second RS's are opposite above the threshold, the analytical continuation of the $G(s)^{\rm FT}$ function in Eq.~\eqref{eq.gftsum} to the second sheet is given by 
\begin{eqnarray}
 G(s)^{\rm FT}_{\rm II} =  G(s)^{\rm FT} - i \frac{q(s)}{4\pi \sqrt{s}} \bigg[ 1 + f(\widetilde{E}_1)  + f(\widetilde{E}_2) \bigg]\,,
\end{eqnarray}
in analogy to the case of Eq.~\eqref{eq.defg2ndrs} at zero temperature. 

Another subtlety in the study of the thermal behaviors of the resonances is the thermal corrections to the pNGBs' masses, which have been the focus of Ref.~\cite{Gu:2018swy}. In this work, we take into account the thermal masses of the $\pi, K, \eta$ and $\eta'$ determined in the previous reference to study their influences on the scalar resonances. The main reason to pay special attention to the thermal masses of the pNGBs is that the threshold effects contained in the $s$-channel unitarity loop functions, which are nonperturbatively resummed, play important roles in the study of the scalar resonances. In contrast, the contributions from the possible crossed channels and the tadpole loops are only perturbatively treated in the present unitarization procedure. Therefore we consider that the changes of threshold effects caused by the shifts of the thermal masses of the pNGBs may play some visible roles in the determination of the thermal properties of the scalar resonances.

\begin{figure}[htbp]
      \centering
   \includegraphics[width=0.72\textwidth,angle=-0]{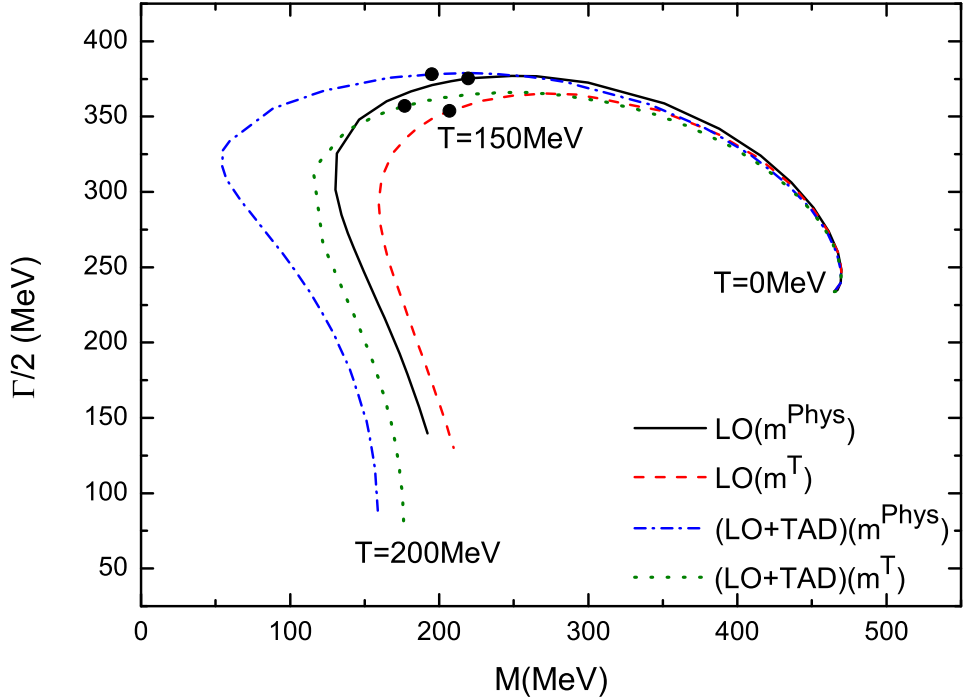} 
  \caption{ The pole trajectories of the $\sigma$ resonance when increasing the temperatures $T$ from 0 to 200~MeV. The black  solid line corresponds to the case of including the LO amplitudes by fixing the physical masses of the $\pi, K, \eta$ and $\eta'$, while the red dashed line denotes the LO result by using the thermal masses of the pNGBs from Ref.~\cite{Gu:2018swy}.  The blue dashed-dotted and green dotted lines stand for the results by including the thermal tadpole corrections with the physical and thermal masses of the pNGBs, respectively. See the text for details.}
   \label{fig.polesigma}
\end{figure} 

\begin{figure}[htbp]
      \centering
   \includegraphics[width=0.72\textwidth,angle=-0]{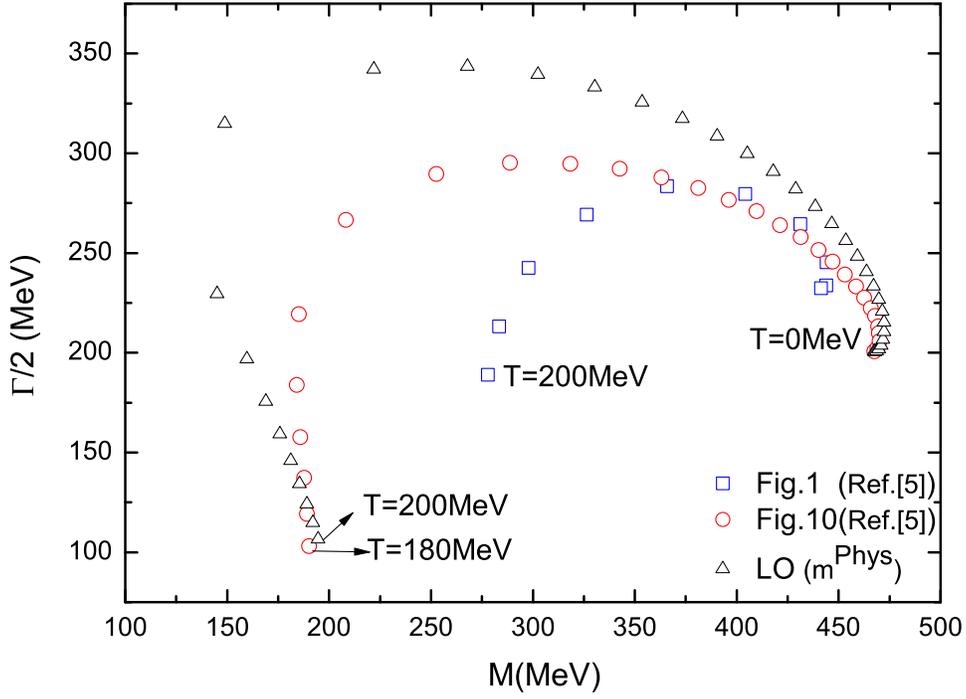} 
  \caption{ Comparisons of the thermal $\sigma$ pole trajectories between ours and those from Ref.~\cite{Cabrera:2008tja}. The black triangles denote our unitarized LO results by using the physical masses of the pNGBs, i.e. the black solid curves shown in Fig.~\ref{fig.polesigma}. The  blue squares stand for the one-loop IAM results, i.e. the left panel of Fig.1 in Ref.~\cite{Cabrera:2008tja} and the red circles correspond to the unitarized results by including the off-shell and tadpole thermal corrections Ref.~\cite{Cabrera:2008tja}, i.e. the left panel of Fig.10 of the latter reference. The blue squares from Ref.~\cite{Cabrera:2008tja} are given by increasing the temperatures in 20~MeV intervals, while the results shown by the black triangles and red circles are obtained by increasing the temperatures in the 5-MeV step.}
   \label{fig.polesigma-compare}
\end{figure}

\begin{figure}[htbp]
      \centering
   \includegraphics[width=0.72\textwidth,angle=-0]{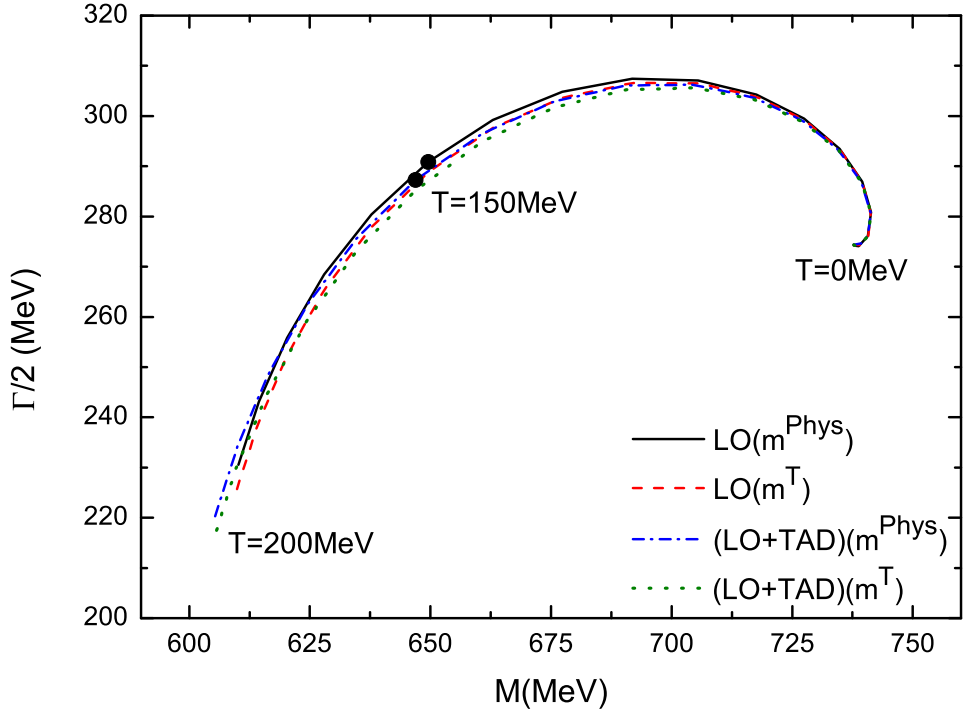} 
  \caption{ The pole trajectories of the $\kappa$ resonance when increasing the temperature $T$ from 0 to 200~MeV. The notations are the same as those in Fig.~\ref{fig.polesigma}. }
   \label{fig.polekappa}
\end{figure} 

The thermal pole trajectories of the $\sigma$ resonance for $0 \leq T \leq 200$~MeV are given in Fig.~\ref{fig.polesigma}. We distinguish the cases by fixing the physical masses of $\pi, K, \eta$ and $\eta'$ and varying their masses at different temperatures according to the results in Ref.~\cite{Gu:2018swy}. We have shifted the masses of the pNGBs at zero temperature in~\cite{Gu:2018swy} to their physical values, in order to match the resonance poles determined in Table~\ref{tab.polezerot}. It turns out that the differences caused by using the different masses of the pNGBs are small, but become visible when the temperatures $T$ are above around 100~MeV. 
Moreover the tadpole loop diagrams in $U(3)$ $\chi$PT, including the 1PI Feynman diagrams shown in Fig.~2(b) of \cite{Guo:2011pa} and the wave function renormalizations, at finite temperatures share the same forms as those at zero temperature, which have been calculated in the former reference. For the sake of completeness, we give the $S$-wave projection of the meson-meson scattering from the tadpole diagrams in the Appendix A. At finite temperatures one only needs to replace the $A_0(m^2)$ loop function by its thermal expression~\cite{LeBellac} 
\begin{eqnarray} \label{eq.a0ft}
A_0(m^2) = - \frac{m^2}{16\pi^2}\ln{\frac{m^2}{\mu^2}} - \int^\infty_0 dp \frac{p^2}{2\pi^2E_p} \frac{1}{e^{\frac{E_p}{T}}-1}\,,
\end{eqnarray}
with $E_p=\sqrt{p^2+m^2}$. Since we are interested in the thermal properties of the scalar resonances, we only include the thermal corrections from the tadpole loop diagrams; that is, we just include the second term in Eq.~\eqref{eq.a0ft} when searching the thermal scalar resonance poles. The blue dashed-dotted and green dotted lines in Fig.~\ref{fig.polesigma} correspond to the results by including the additional thermal corrections from the tadpole loop diagrams. Apparently the inclusion of the additional thermal corrections from the tadpole loop diagrams only slightly affects the thermal trajectories of the $\sigma$ resonance. The most important lesson we learn from Fig.~\ref{fig.polesigma} is that the mass of the $\sigma$ significantly decreases even below the $\pi\pi$ threshold when increasing the temperatures $T$ up to 200~MeV. This seems consistent with the requirement of the chiral symmetry restoration~\cite{Hatsuda:1985eb,Cortes:2015emo}. However the width of the $\sigma$ is still quite large even when $T$ reaches around 200~MeV. 

The comparisons of the thermal $\sigma$ pole trajectories from our study and those of Ref.~\cite{Cabrera:2008tja} are explicitly shown in Fig.~\ref{fig.polesigma-compare}. Although it is quantitatively different from 
the full one-loop IAM study, the unitarization of the LO $U(3)$ $\chi$PT indeed gives similar result as the single-channel study by including both the off-shell and tadpole effects\footnote{We have exactly reproduced the thermal $\sigma$ poles in the left panel of Fig.(10) in Ref.~\cite{Cabrera:2008tja}, by correcting a typesetting typo in Eq.(13) of that reference, i.e. to multiply 1/2 in the first two terms in Eq.(13). } in Ref.~\cite{Cabrera:2008tja}. We verify that by neglecting the off-shell terms in Eq.~(12) of the latter reference the results are only slightly changed. Therefore the on-shell prescription in Eq.~\eqref{eq.utft} by only including the contributions from the $\chi$PT amplitudes to the $\mK(s)$ will be used throughout this work. The qualitative similarities of the curves in Fig.~\ref{fig.polesigma-compare} indicate that the thermal corrections to the $G(s)$ function include the important part of the finite-temperature effects in the study of the scalar resonances.

Since the most important channel for the $\sigma$ resonance is the $\pi\pi$, there is no problem of the Landau cut, due to the equal-mass feature in this scattering. The only unequal-mass process of the $U(3)$ coupled-channel scattering in the isoscalar scalar case is the $\eta\eta'$ one, which thermal Landau cut extends up to $(m_{\eta'}-m_{\eta})^2$ around (400~MeV)$^2$ in the real $s$ axis. In order to check the possible influence of this Landau cut, we have performed a three-channel study, i.e. $\pi\pi, K\bar{K}$ and $\eta\eta$, where the Landau cuts will not be problems, due to the equal-mass features. It turns out that the resulting curves are almost indistinguishable from the five-channel scattering by including $\eta\eta'$ and $\eta'\eta'$. This in turn implies that the $\eta\eta'$ channel, including its Landau cut, plays unimportant roles in the determination of the thermal $\sigma$ poles.

\begin{figure}[htbp]
   \centering
   \begin{minipage}[t]{0.49\textwidth}
      \centering
   \includegraphics[width=0.99\textwidth,angle=-0]{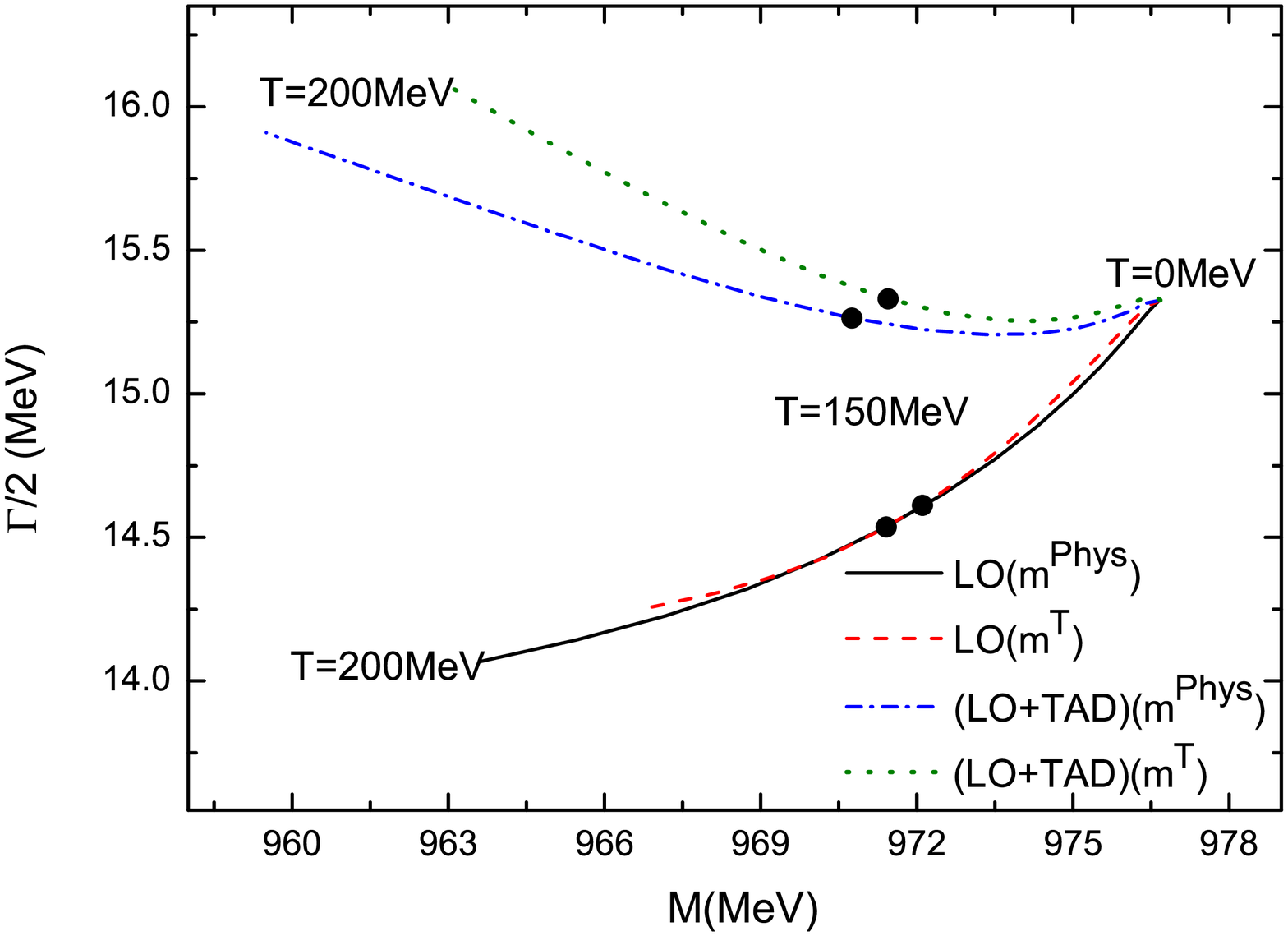} 
   \end{minipage}
      \begin{minipage}[t]{0.49\textwidth}
         \centering
   \includegraphics[width=0.99\textwidth,angle=-0]{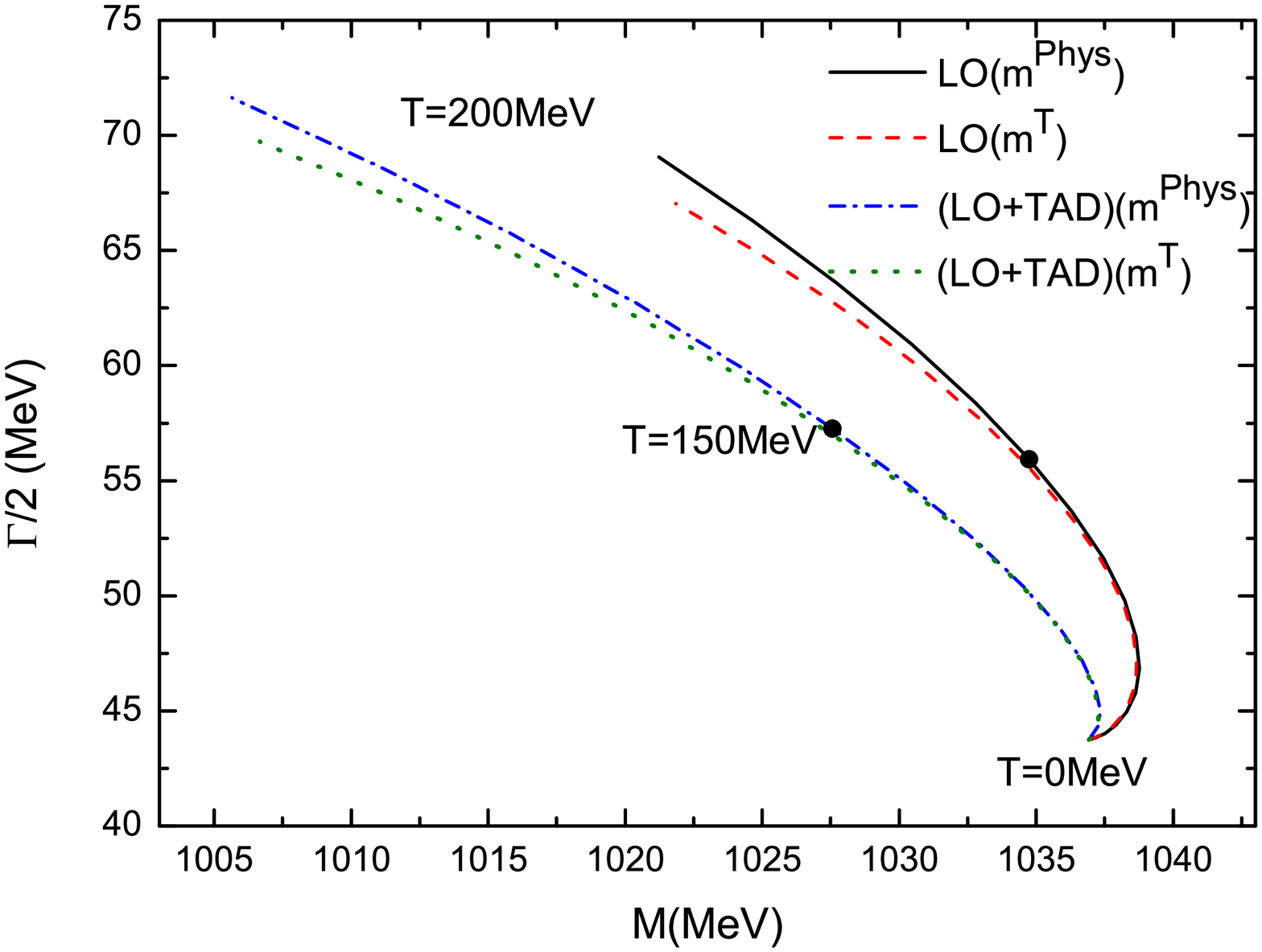} 
   \end{minipage}
  \caption{ The thermal behaviors of the poles of the $f_0(980)$ (left) and $a_0(980)$ (right).  The notations are the same as those in  Fig.~\ref{fig.polesigma}. }
   \label{fig.polef0a0}
\end{figure}

The pole trajectories of the $\kappa$ resonance with varying temperatures are given in Fig.~\ref{fig.polekappa}. Clearly the thermal behaviors of the $\kappa$ pole share similar trends as the $\sigma$, with a significant decrease of the mass and moderate change of the width, when increasing the temperatures up to 200~MeV. In Fig.~\ref{fig.polef0a0}, we show the pole trajectories of the $f_0(980)$ and $a_0(980)$ resonances. Unlike the $\sigma$ and $\kappa$, the poles of the $f_0(980)$ and $a_0(980)$ are insensitive to the changes of the temperatures, and both the masses and widths are only slightly changed. E.g., the masses of the $\sigma$ and $\kappa$ decrease around 300~MeV and 100~MeV, respectively, when varying the temperatures from 0 to 200~MeV. In contrast, the masses of the $f_0(980)$ and $a_0(980)$ only decrease around 15~MeV and 30~MeV, in order. Comparing with the black solid and red dashed lines, and the blue dashed-dotted and green dotted lines in Figs.~\ref{fig.polesigma}, \ref{fig.polekappa} and \ref{fig.polef0a0}, we can conclude that the thermal corrections to the masses of the $\pi, K, \eta$ and $\eta'$ marginally affect the properties of the scalar resonances $\sigma, f_0(980), \kappa$ and $a_0(980)$. The effects of the thermal corrections from the tadpole loop diagrams are also small for all the cases shown in Figs.~\ref{fig.polesigma}, \ref{fig.polekappa} and \ref{fig.polef0a0}. 

Regarding the thermal Landau cut in the $K\pi$ scattering, which is the most important channel for the $\kappa$, it extends to the positive real $s$ axis up to $(m_K-m_\pi)^2$ around (360~MeV)$^2$, which is clearly distant from the thermal $\kappa$ poles shown in Fig.~\ref{fig.polekappa}. For the inelastic $K\eta$ and $K\eta'$ channels, their thermal Landau cuts end around $50^2$ and $460^2$~MeV$^2$ in the positive real $s$ axis, which are also far away from the $\kappa$ poles. The most important three channels for $f_0(980)$ are $\pi\pi, K\bar{K}$ and $\eta\eta$ and there are no issues of the thermal Landau cuts, due to their equal-mass features. While for $a_0(980)$, the most important two channels are the $\pi\eta$ and $K\bar{K}$. The thermal Landau cut in the $\pi\eta$ scattering extends to the positive real $s$ axis up to $(m_\eta-m_\pi)^2$ around (410~MeV)$^2$, which is distant from the $a_0(980)$ poles shown in Fig.~\ref{fig.polef0a0}. The Landau cut in the $\pi\eta'$ channel extends to the real positive $s$ axis up to $(m_\eta'-m_\pi)^2$ around (820~MeV)$^2$. We have explicitly verified that the resulting thermal $a_0(980)$ poles from the $\pi\eta$ and $K\bar{K}$ coupled-channel scattering are almost identical to the three coupled-channel scattering by including the $\pi\eta'$ channel, which implies that the $\pi\eta'$ channel, including its Landau cut, plays a negligible role in the determination of the $a_0(980)$.

\section{Summary and conclusions}~\label{sec.conclusion}

In this work the light-flavor QCD scalar resonances $\sigma$, $\kappa$, $f_0(980)$ and $a_0(980)$ are studied in the framework of the unitarized $U(3)$ chiral perturbation theory. Special attention is paid to their thermal properties, including the trajectories of their resonance pole positions with varying temperatures. Different from the works that only study on the thermal masses of the scalar resonances, e.g. Refs.~\cite{Jiang:2012wm,Jiang:2015xqz}, we first fix the unknown parameters by fitting the experimental and lattice data of the meson-meson scattering, which enables us to obtain reliable resonance properties at zero temperature, including both the masses and widths. The finite-temperature effects are included through the unitarization procedure and the tadpole loop diagrams. It turns out that the thermal corrections from the unitarity loop functions are the most important parts, while the tadpole diagrams only play minor roles. The key merit of our approach is that we do not need to introduce any new parameter in the study of the thermal behaviors of the scalar resonances, once they are determined in the vacuum. 

The $\sigma$ pole trajectories of the present study are qualitatively similar to those in the previous  works~\cite{GomezNicola:2002tn,FernandezFraile:2007fv}, which are obtained by including the complete one-loop thermal corrections to the $\pi\pi$ scattering. This validates the current approach to include the finite-temperature effects via the unitarization procedure. Our results show that the masses of the $\sigma$ and $\kappa$ significantly decrease when increasing the temperatures up to 200~MeV, while their widths are moderately changed and remain large. In contrast, both the masses and widths of the $f_0(980)$ and $a_0(980)$ are insensitive to the temperatures. The present formalism provides an efficient and straightforward way to study thermal behaviors of other types of resonances. We expect to apply this approach to other systems at finite temperatures in the future.

\section*{Acknowledgements}
We thank Yu~Jia, J.~Ruiz de Elvira and Mei-Zhu~Yan for useful comments and discussions. 
This work is supported in part by the National Natural Science Foundation of China (NSFC) under Grants No.~11575052, No.~11975090 and the Natural Science Foundation of Hebei Province under Contract No.~A2015205205.

\section*{Appendix A: S-wave meson-meson scattering amplitudes}\label{sec.appendixA}

First we give the leading-order $S$-wave meson-meson scattering amplitudes $T_{IJ}(s)$ with definite isopsin number in the $U(3)$ $\chi$PT~\cite{Guo:2011pa}. There are five coupled channels for the $(I,J)=(0,0)$ case and they read 
\begin{eqnarray}\label{eq.pw00}
T_{00}^{\pi\pi\to\pi\pi}(s) &=&\dfrac{2s-m_\pi^2}{2F_\pi^2} \,, \nonumber \\
T_{00}^{\pi\pi\to K\bar{K}}(s) &=& \dfrac{\sqrt{3}s}{4F_\pi^2} \,, \nonumber \\
T_{00}^{\pi\pi\to\eta\eta}(s) &=& \dfrac{-\sqrt{3}m_\pi^2(c_{\theta}-\sqrt{2}s_{\theta})^2}{6F_\pi^2}\,, \nonumber \\
T_{00}^{\pi\pi\to\eta\eta'}(s) &=& \dfrac{-\sqrt{3}m_\pi^2(\sqrt{2}c^2_{\theta}-c_{\theta}s_{\theta}-\sqrt{2}s^2_{\theta})}{3\sqrt{2}F_\pi^2}\,, \nonumber \\
T_{00}^{\pi\pi\to\eta'\eta'}(s) &=& \dfrac{-\sqrt{3}m_\pi^2(\sqrt{2}c_{\theta}+s_{\theta})^2}{6F_\pi^2}\,, \nonumber \\
T_{00}^{K\bar{K}\to K\bar{K}}(s) &=&  \dfrac{3s}{4F_\pi^2} \,, \nonumber \\
T_{00}^{K\bar{K}\to\eta\eta}(s) &=& \dfrac{-[(-6m_\eta^2-2m_\pi^2+9s)c^2_{\theta}+4\sqrt{2}(2m_K^2-m_\pi^2)c_{\theta}s_{\theta}+8m_K^2s^2_{\theta}]}{12F_\pi^2}\,, \nonumber \\
T_{00}^{K\bar{K}\to\eta\eta'}(s) &=& \dfrac{-2\sqrt{2}c^2_{\theta}(m_\pi^2-2m_K^2)+c_{\theta}s_{\theta}(3m_\eta^2+3m_{\eta^\prime}^2+8m_K^2+2m_\pi^2-9s)-2\sqrt2 s^2_{\theta}(2m_K^2-m_\pi^2)}{6\sqrt{2}F_\pi^2}\,, \nonumber \\
T_{00}^{K\bar{K}\to\eta'\eta'}(s) &=& \dfrac{-8c^2_{\theta}m_K^2+4\sqrt{2}c_{\theta}s_{\theta}(2m_K^2-m_\pi^2)-s^2_{\theta}(9s-6m_{\eta^\prime}^2-2m_\pi^2)}{12F_\pi^2}\,, \nonumber \\
T_{00}^{\eta\eta\to\eta\eta}(s) &=& \dfrac{[c^4_{\theta}(16m_K^2-7m_\pi^2)+4\sqrt{2}c^3_{\theta}s_{\theta}(8m_K^2-5m_\pi^2)+12c^2_{\theta}s^2_{\theta}(4m_K^2-m_\pi^2)]}{18F_\pi^2} \nonumber \\
&&+\dfrac{[16\sqrt{2}c_{\theta}s^3_{\theta} (m_K^2-m_\pi^2)+2s^4_{\theta}(2m_K^2+m_\pi^2)]}{18F_\pi^2} \,, \nonumber \\
T_{00}^{\eta\eta\to\eta\eta'}(s) &=& \dfrac{[\sqrt{2}c^4_{\theta}(-8m_K^2+5m_\pi^2)-c^3_{\theta}s_{\theta}(8m_K^2+m_\pi^2)+3\sqrt{2}c^2_{\theta}s^2_{\theta}(4m_K^2-m_\pi^2)]}{9\sqrt{2}F_\pi^2} \nonumber \\
&&+\dfrac{[4c_{\theta}s^3_{\theta} (5m_K^2-2m_\pi^2)+4\sqrt{2}s^4_{\theta}(m_K^2-m_\pi^2)]}{9\sqrt{2}F_\pi^2} \,, \nonumber \\
T_{00}^{\eta\eta\to\eta'\eta'}(s) &=& \dfrac{(4m_K^2-m_\pi^2)(2c^4_{\theta}-2\sqrt{2}c^3_{\theta}s_{\theta}-3c^2_{\theta}s^2_{\theta}+2\sqrt{2}c_{\theta}s^3_{\theta}+2s^4_{\theta})}{18F_\pi^2} \,, \nonumber \\
T_{00}^{\eta\eta'\to\eta\eta'}(s) &=& \dfrac{(4m_K^2-m_\pi^2)(2c^4_{\theta}-2\sqrt{2}c^3_{\theta}s_{\theta}-3c^2_{\theta}s^2_{\theta}+2\sqrt{2}c_{\theta}s^3_{\theta}+2s^4_{\theta})}{9F_\pi^2} \,, \nonumber \\
T_{00}^{\eta\eta'\to\eta'\eta'}(s) &=& \dfrac{[4\sqrt{2}c^4_{\theta}(-m_K^2+m_\pi^2)+4c^3_{\theta}s_{\theta}(5m_K^2-2m_\pi^2)+3\sqrt{2}c^2_{\theta}s^2_{\theta}(-4m_K^2+m_\pi^2)]}{9\sqrt{2}F_\pi^2} \nonumber \\
 &&-\dfrac{[c_{\theta}s^3_{\theta} (8m_K^2+m_\pi^2)-\sqrt{2}s^4_{\theta}(8m_K^2-5m_\pi^2)]}{9\sqrt{2}F_\pi^2}\,, \nonumber \\
T_{00}^{\eta'\eta'\to\eta'\eta'}(s) &=& \dfrac{[2c^4_{\theta}(2m_K^2+m_\pi^2)-16\sqrt{2}c^3_{\theta}s_{\theta}(m_K^2-m_\pi^2)+12c^2_{\theta}s^2_{\theta}(4m_K^2-m_\pi^2)]}{18F_\pi^2}\nonumber  \\
 &&+ \dfrac{-4\sqrt{2}c_{\theta}s^3_{\theta} (8m_K^2-5m_\pi^2)+s^4_{\theta}(16m_K^2-7m_\pi^2)}{18F_\pi^2}\,,
\end{eqnarray}
with $s_\theta=\sin\theta, c_\theta=\cos\theta$ and $\theta$ the LO $\eta$-$\eta'$ mixing angle given in Eq.~\eqref{eq.loangle}. 
 
For the $(I,J)=(1,0)$ case, there are three coupled channels and the explicit results are 
\begin{eqnarray}\label{eq.pw10}
T_{10}^{\pi\eta\to\pi\eta}(s) &=& \dfrac{(c_\theta-\sqrt2 s_\theta)^2m_\pi^2}{3 F_\pi^2}\,,\nonumber \\
T_{10}^{\pi\eta\to K\bar{K}}(s) &=& \dfrac{c_\theta(3m_\eta^2+8 m_K^2 + m_\pi^2-9s)+2\sqrt{2}s_\theta(2m_K^2+m_\pi^2) }{6\sqrt{6} F_\pi^2}\,,\nonumber \\
T_{10}^{\pi\eta\to\pi\eta'}(s) &=& \dfrac{(\sqrt{2}c_\theta^2-c_\theta s_\theta - \sqrt2 s_\theta^2)m_\pi^2}{3 F_\pi^2}\,,\nonumber \\
T_{10}^{K\bar{K}\to K\bar{K}}(s) &=& \dfrac{s}{4 F_\pi^2}\,,\nonumber \\
T_{10}^{K\bar{K}\to \pi\eta'}(s) &=& \dfrac{s_\theta(3m_{\eta'}^2+8 m_K^2 + m_\pi^2-9s) - 2\sqrt{2}c_\theta(2m_K^2+m_\pi^2) }{6\sqrt{6} F_\pi^2}\,,\nonumber \\
T_{10}^{\pi\eta'\to\pi\eta'}(s) &=& \dfrac{(\sqrt2 c_\theta + s_\theta)^2m_\pi^2}{3 F_\pi^2}\,.
\end{eqnarray}

For the $(I,J)=(\frac{1}{2},0)$ case, there are three coupled channels and their amplitudes take the form 
\begin{eqnarray}\label{eq.pw1d20}
T_{\frac{1}{2}0}^{K\pi\to K\pi}(s) &=&  \dfrac{-3(m_K^2-m_\pi^2)^2-2(m_K^2+m_\pi^2)s+5s^2}{8sF_\pi^2}\,,\nonumber \\
T_{\frac{1}{2}0}^{K\pi\to K\eta}(s) &=& \frac{c_{\theta}\big[6m_\eta^2-20m_K^2+2m_\pi^2+9\frac{(-m_\eta^2+m_K^2+s)(m_K^2-m_\pi^2+s)}{s}\big]+4\sqrt{2}(2m_K^2+m_\pi^2)s_{\theta}}{24F_\pi^2} \,,\nonumber \\
T_{\frac{1}{2}0}^{K\pi\to K\eta'}(s) &=& \frac{-4\sqrt{2}c_{\theta}(2m_K^2+m_\pi^2)+\big[6m_{\eta'}^2-20m_K^2+2m_\pi^2+9\frac{(-m_{\eta'}^2+m_K^2+s)(m_K^2-m_\pi^2+s)}{s}\big]s_{\theta}}{24F_\pi^2} \,,\nonumber \\
T_{\frac{1}{2}0}^{K\eta\to K\eta}(s) &=& \frac{1}{{24F_\pi^2s}}\big[c_{\theta}^2(-9m_\eta^4-9m_K^4+18m_K^2s-4m_\pi^2s-9s^2+ 18m_\eta^2 m_K^2+ 6 m_\eta^2 s ) \nonumber \\ &&
+8\sqrt{2}c_{\theta}s_{\theta}\,s(2m_K^2-m_\pi^2)+16s\,m_K^2 s_{\theta}^2 \big] \,,\nonumber \\
T_{\frac{1}{2}0}^{K\eta\to K\eta'}(s) &=&   \dfrac{4\sqrt{2}c_{\theta}^2(-2m_K^2+m_\pi^2)-c_{\theta}s_{\theta} \big[6m_\eta^2+6m_{\eta'}^2-20m_K^2+4m_\pi^2
 +9\frac{(-m_\eta^2+m_K^2+s)(-m_{\eta'}^2+m_K^2+s)}{s} \big]}{24F_\pi^2} \nonumber \\ &&
+\frac{4\sqrt{2}(2m_K^2-m_\pi^2)s_{\theta}^2}{24F_\pi^2} \,,\nonumber \\
T_{\frac{1}{2}0}^{K\eta'\to K\eta'}(s) &=&  \dfrac{1}{{24F_\pi^2s}} \big[ 16c_{\theta}^2m_K^2s+8\sqrt{2}c_{\theta}s_{\theta}(-2m_K^2+m_\pi^2)s \nonumber \\ && +s_{\theta}^2 (-9m_{\eta'}^4-9m_K^4+18m_K^2s-4m_\pi^2s-9s^2+18m_{\eta'}^2m_K^2 + 6m_{\eta'}^2s)\big] \,.
\end{eqnarray}

In addition, we also consider the tadpole loop diagrams, which consist of the 1PI tadpole loops and the wave function renormalizations. We mention that the chiral tadpole loops of the scattering processes at finite temperatures share the same structures as those appearing in the zero temperature case. To include the thermal corrections of the tadpole loops, one only needs to replace the $A_0(m^2)$ function by its finite-temperature counterpart, see Eq.~\eqref{eq.a0ft}. The $S$-wave projections of the tadpole loop amplitudes $T_{IJ,TAD}(s)$ with definite isopsin are given below. For the $(I,J)=(0,0)$ case they read
\begin{eqnarray} \label{pw00tad}
&&T_{00,TAD}^{\pi\pi\to\pi\pi}(s)=\frac{5 m_{\pi }^2 \left(2 c_{\theta }^2+2 \sqrt{2} c_{\theta } s_{\theta }+s_{\theta }^2\right) A_0(m_{\eta '}^2)}{36 F_{\pi }^4}+\frac{5 m_{\pi }^2 A_0(m_{\eta }^2)
   \left(c_{\theta }^2-2 \sqrt{2} c_{\theta } s_{\theta }+2 s_{\theta }^2\right)}{36 F_{\pi }^4}\nonumber\\&&-\frac{\left(8 m_{\pi }^2+4 s\right) A_0(m_K^2)}{36 F_{\pi
   }^4}-\frac{A_0(m_{\pi }^2) \left(51 m_{\pi }^2+8 s\right)}{36 F_{\pi }^4} \,,\nonumber
\end{eqnarray}
\begin{eqnarray}
&&T_{00,TAD}^{\pi\pi\to K\bar{K}}(s) =-\frac{A_0(m_{\eta '}^2)}{240 \sqrt{3} F_{\pi }^4} \bigg[-40 c_{\theta }^2 \left(m_K^2+m_{\pi }^2\right)+4 \sqrt{2} c_{\theta } \left(2 m_K^2-7 m_{\pi }^2\right) s_{\theta }\nonumber\\&&+s_{\theta }^2 \left(64
   m_K^2+12 m_{\eta '}^2+4 m_{\pi }^2-15 s\right)\bigg] -\frac{A_0(m_K^2)
   \left(16 m_K^2+4 m_{\pi }^2+s\right)}{24 \sqrt{3} F_{\pi }^4}-\frac{A_0(m_{\pi }^2) \left(40 m_{\pi }^2+13 s\right)}{48 \sqrt{3} F_{\pi }^4} \nonumber\\&& -\frac{A_0(m_{\eta }^2)}{240 \sqrt{3} F_{\pi }^4} \bigg[c_{\theta }^2 \left(64 m_K^2+12 m_{\eta }^2+4 m_{\pi }^2-15
   s\right)+4 \sqrt{2} c_{\theta } \left(7 m_{\pi }^2-2 m_K^2\right) s_{\theta }-40 \left(m_K^2+m_{\pi }^2\right) s_{\theta }^2\bigg] \,, \nonumber
\end{eqnarray}
\begin{eqnarray}
&& T_{00,TAD}^{\pi\pi\to\eta\eta}(s)= -\frac{A_0(m_K^2)}{30 \sqrt{3} F_{\pi }^4} \bigg[-c_{\theta }^2 \left(16 m_K^2+3 m_{\eta }^2+30 m_{\pi }^2  s_{\theta }^2+6m_{\pi}^2\right)+\sqrt{2} c_{\theta } \left(2 m_K^2+3 m_{\pi }^2\right)
   s_{\theta }\nonumber\\&& -15 m_{\pi }^2 c_{\theta }^4+30 \sqrt{2} m_{\pi }^2 c_{\theta }^3 s_{\theta }+10 m_K^2 s_{\theta }^2\bigg] -\frac{m_{\pi }^2 \left(2 c_{\theta }^4-2
   \sqrt{2} c_{\theta }^3 s_{\theta }-3 c_{\theta }^2 s_{\theta }^2+2 \sqrt{2} c_{\theta } s_{\theta }^3+2 s_{\theta }^4\right) A_0(m_{\eta '}^2)}{12 \sqrt{3} F_{\pi
   }^4}\nonumber\\&&-\frac{m_{\pi }^2 A_0(m_{\eta }^2) \left(c_{\theta }^4-4 \sqrt{2} c_{\theta }^3 s_{\theta }+12 c_{\theta }^2 s_{\theta }^2-8 \sqrt{2} c_{\theta } s_{\theta }^3+4 s_{\theta
   }^4\right)}{12 \sqrt{3} F_{\pi }^4} -\frac{m_{\pi }^2 A_0(m_{\pi }^2) \left(c_{\theta }^2-2 \sqrt{2} c_{\theta } s_{\theta }+2 s_{\theta }^2\right)}{12 \sqrt{3} F_{\pi }^4}\,, \nonumber
   \end{eqnarray}
\begin{eqnarray}
&&T_{00,TAD}^{\pi\pi\to\eta\eta'}(s)= -\frac{A_0(m_K^2)}{30 \sqrt{6} F_{\pi }^4} \bigg[-\sqrt{2} c_{\theta }^2 \left(2 m_K^2+3 m_{\pi }^2\right)-c_{\theta } s_{\theta } \left(52 m_K^2+3 m_{\eta '}^2+3 m_{\eta }^2-15 m_{\pi }^2 s_{\theta
   }^2+12 m_{\pi }^2\right)\nonumber\\&&-15 \sqrt{2} m_{\pi }^2 c_{\theta }^4+15 m_{\pi }^2 c_{\theta }^3 s_{\theta }+\sqrt{2} s_{\theta }^2 \left(2 m_K^2+15 m_{\pi }^2  s_{\theta
   }^2+3m_{\pi}^2\right)\bigg]-\frac{m_{\pi }^2 A_0(m_{\pi }^2)
   \left(\sqrt{2} c_{\theta }^2-c_{\theta } s_{\theta }-\sqrt{2} s_{\theta }^2\right)}{6 \sqrt{6} F_{\pi }^4} \nonumber\\&&-\frac{m_{\pi }^2 \left(2 \sqrt{2} c_{\theta }^4+2 c_{\theta }^3 s_{\theta }-3 \sqrt{2} c_{\theta }^2 s_{\theta }^2-5 c_{\theta }
   s_{\theta }^3-\sqrt{2} s_{\theta }^4\right) A_0(m_{\eta '}^2)}{6 \sqrt{6} F_{\pi }^4}\nonumber\\&&-\frac{m_{\pi }^2 A_0(m_{\eta }^2) \left(\sqrt{2} c_{\theta }^4-5 c_{\theta }^3
   s_{\theta }+3 \sqrt{2} c_{\theta }^2 s_{\theta }^2+2 c_{\theta } s_{\theta }^3-2 \sqrt{2} s_{\theta }^4\right)}{6 \sqrt{6} F_{\pi }^4}\,, \nonumber
\end{eqnarray}
\begin{eqnarray}
&&T_{00,TAD}^{\pi\pi\to\eta'\eta'}(s)=-\frac{A_0(m_K^2)}{30 \sqrt{3} F_{\pi }^4} \bigg[10 c_{\theta }^2 \left(m_K^2-3 m_{\pi }^2 s_{\theta }^2\right)-\sqrt{2} c_{\theta } s_{\theta } \left(2 m_K^2+30 m_{\pi }^2  s_{\theta
   }^2+3m_{\pi}^2\right)\nonumber\\&&-s_{\theta }^2 \left(16 m_K^2+3 m_{\eta '}^2+15 m_{\pi }^2  s_{\theta }^2+6m_{\pi}^2\right)\bigg] -\frac{m_{\pi }^2 A_0(m_{\eta }^2) \left(2 c_{\theta }^4-2 \sqrt{2} c_{\theta }^3 s_{\theta }-3 c_{\theta }^2 s_{\theta }^2+2 \sqrt{2} c_{\theta } s_{\theta }^3+2
   s_{\theta }^4\right)}{12 \sqrt{3} F_{\pi }^4}  \nonumber\\&&-\frac{m_{\pi }^2 \left(4 c_{\theta
   }^4+8 \sqrt{2} c_{\theta }^3 s_{\theta }+12 c_{\theta }^2 s_{\theta }^2+4 \sqrt{2} c_{\theta } s_{\theta }^3+s_{\theta }^4\right) A_0(m_{\eta '}^2)}{12 \sqrt{3} F_{\pi
   }^4} -\frac{m_{\pi }^2 A_0(m_{\pi }^2) \left(2 c_{\theta }^2+2 \sqrt{2} c_{\theta } s_{\theta }+s_{\theta }^2\right)}{12 \sqrt{3}
   F_{\pi }^4} \,, \nonumber
      \end{eqnarray}
\begin{eqnarray}
&&T_{00,TAD}^{K\bar{K}\to K\bar{K}}(s)=  \frac{A_0(m_{\eta '}^2) \bigg[20 c_{\theta }^2 m_K^2+2 \sqrt{2} c_{\theta } \left(3 m_{\pi }^2-8 m_K^2\right) s_{\theta }-s_{\theta }^2 \left(8 m_K^2+24 m_{\eta '}^2+3 m_{\pi
   }^2+15 s\right)\bigg]}{60 F_{\pi }^4}\nonumber\\&&+\frac{A_0(m_{\eta }^2) \bigg[-c_{\theta }^2 \left(8 m_K^2+24 m_{\eta }^2+3 m_{\pi }^2+15 s\right)-2 \sqrt{2} c_{\theta }
   \left(3 m_{\pi }^2-8 m_K^2\right) s_{\theta }+20 m_K^2 s_{\theta }^2\bigg]}{60 F_{\pi }^4}\nonumber\\&&-\frac{m_K^2 A_0(m_K^2)}{F_{\pi }^4}-\frac{A_0(m_{\pi }^2) \left(4
   m_K^2+m_{\pi }^2\right)}{12 F_{\pi }^4} \,,\nonumber
\end{eqnarray}
\begin{eqnarray}
&&  T_{00,TAD}^{K\bar{K}\to\eta\eta}(s) =\frac{A_0(m_{\eta '}^2)}{720 F_{\pi }^4} \bigg[-40 c_{\theta }^4 \left(3 m_K^2-m_{\pi }^2\right)-3 c_{\theta }^2 s_{\theta }^2 \left(-128 m_K^2-12 m_{\eta '}^2+18 m_{\eta }^2+42 m_{\pi }^2+15
   s\right)\nonumber\\&&+4 \sqrt{2} c_{\theta }^3 \left(6 m_K^2-m_{\pi }^2\right) s_{\theta }+8 \sqrt{2} c_{\theta } \left(12 m_K^2-7 m_{\pi }^2\right) s_{\theta }^3+40 m_{\pi }^2 s_{\theta
   }^4\bigg]\nonumber\\&&-\frac{A_0(m_K^2)}{120 F_{\pi }^4} \bigg[-c_{\theta }^2 \left(80 m_K^2  s_{\theta }^2+32m_K^2+66 m_{\eta }^2+2 m_{\pi }^2-45 s\right)+40 \sqrt{2} c_{\theta
   }^3 \left(m_{\pi }^2-2 m_K^2\right) s_{\theta }\nonumber\\&&+4 \sqrt{2} c_{\theta } \left(6 m_K^2-m_{\pi }^2\right) s_{\theta }+10 c_{\theta }^4 \left(6 m_{\eta }^2+2 m_{\pi }^2-9 s\right)+40 m_K^2
   s_{\theta }^2\bigg]\nonumber\\&&-\frac{A_0(m_{\pi }^2) \left[c_{\theta }^2 \left(-64 m_K^2+18 m_{\eta }^2+6 m_{\pi }^2+15 s\right)-8 \sqrt{2} c_{\theta } \left(4
   m_K^2+m_{\pi }^2\right) s_{\theta }+40 m_{\pi }^2 s_{\theta }^2\right]}{240 F_{\pi }^4}\nonumber\\&&+\frac{A_0(m_{\eta }^2)}{720 F_{\pi }^4} \bigg[c_{\theta }^4 \left(64 m_K^2-18 m_{\eta }^2+34 m_{\pi
   }^2-45 s\right)-4 \sqrt{2} c_{\theta }^3 \left(62 m_K^2-27 m_{\pi }^2\right) s_{\theta }\nonumber\\&&-120 c_{\theta }^2 \left(5 m_K^2-2 m_{\pi }^2\right) s_{\theta }^2-160 \sqrt{2} c_{\theta }
   \left(2 m_K^2-m_{\pi }^2\right) s_{\theta }^3-160 m_K^2 s_{\theta }^4\bigg]\,, \nonumber
\end{eqnarray}
\begin{eqnarray}
&&T_{00,TAD}^{K\bar{K}\to\eta\eta'}(s)= -\frac{A_0(m_K^2)}{60 \sqrt{2} F_{\pi }^4} \Big\{10 \sqrt{2} c_{\theta }^4 \left(2 m_K^2-m_{\pi }^2\right)+2 \sqrt{2} c_{\theta }^2 \left(m_{\pi }^2-6 m_K^2\right)\nonumber\\&&+5 c_{\theta }^3 s_{\theta } \left(8
   m_K^2+3 m_{\eta '}^2+3 m_{\eta }^2+2 m_{\pi }^2-9 s\right)\nonumber\\&&+c_{\theta } s_{\theta } \big[40 m_K^2 s_{\theta }^2-72 m_K^2+3 \left(5 s_{\theta }^2-11\right) m_{\eta '}^2+3 m_{\eta }^2
   \left(5 s_{\theta }^2-11\right)+10 m_{\pi }^2 s_{\theta }^2-2 m_{\pi }^2\big]\nonumber\\&&+c_{\theta } s_{\theta } \left(-45 s s_{\theta }^2+45 s\right)-2 \sqrt{2} s_{\theta }^2 \left(10 m_K^2  s_{\theta }^2-6m_K^2+m_{\pi }^2
   -5 m_{\pi}^2s_{\theta }^2\right)\Big\}\nonumber\\&&-\frac{A_0(m_{\pi }^2)}{120 \sqrt{2} F_{\pi
   }^4} \bigg[4 \sqrt{2} c_{\theta }^2 \left(4 m_K^2+m_{\pi }^2\right)+c_{\theta }
   s_{\theta } \left(-64 m_K^2+9 m_{\eta '}^2+9 m_{\eta }^2-34 m_{\pi }^2+15 s\right)\nonumber\\&&-4 \sqrt{2} \left(4 m_K^2+m_{\pi }^2\right) s_{\theta }^2\bigg]\nonumber\\&&-\frac{A_0(m_{\eta }^2)}{360 \sqrt{2} F_{\pi }^4} \bigg[-4 \sqrt{2} c_{\theta }^4 \left(8 m_K^2-3 m_{\pi }^2\right)+c_{\theta }^3 s_{\theta } \left(-304 m_K^2+27 m_{\eta '}^2-9 m_{\eta }^2+86
   m_{\pi }^2+45 s\right)\nonumber\\&&-24 \sqrt{2} c_{\theta }^2 \left(m_K^2-m_{\pi }^2\right) s_{\theta }^2+40 c_{\theta } \left(5 m_K^2-3 m_{\pi }^2\right) s_{\theta }^3+40 \sqrt{2} \left(2
   m_K^2-m_{\pi }^2\right) s_{\theta }^4\bigg]\nonumber\\&&-\frac{A_0(m_{\eta '}^2)}{360 \sqrt{2} F_{\pi }^4} \bigg[40 \sqrt{2} c_{\theta }^4 \left(m_{\pi }^2-2 m_K^2\right)+c_{\theta }
   s_{\theta }^3 \left(-304 m_K^2-9 m_{\eta '}^2+27 m_{\eta }^2+86 m_{\pi }^2+45 s\right)\nonumber\\&&+40 c_{\theta }^3 \left(5 m_K^2-3 m_{\pi }^2\right) s_{\theta }+24 \sqrt{2} c_{\theta }^2
   \left(m_K^2-m_{\pi }^2\right) s_{\theta }^2+4 \sqrt{2} \left(8 m_K^2-3 m_{\pi }^2\right) s_{\theta }^4\bigg]\,, \nonumber
   \end{eqnarray}
\begin{eqnarray}
&&T_{00,TAD}^{K\bar{K}\to\eta'\eta'}(s)= \frac{A_0(m_K^2)}{120
   F_{\pi }^4} \bigg[40 c_{\theta }^2 m_K^2 \left(2 s_{\theta }^2-1\right)-4 \sqrt{2} c_{\theta } s_{\theta } \left(20m_K^2  s_{\theta }^2-6m_K^2+m_{\pi }^2 -10
   m_{\pi}^2s_{\theta }^2\right)\nonumber\\&&+s_{\theta }^2 \left(32 m_K^2+66 m_{\eta '}^2-60 s_{\theta }^2 m_{\eta '}^2-20 m_{\pi }^2 s_{\theta }^2+2 m_{\pi }^2+90 s s_{\theta }^2-45 s\right)\bigg]\nonumber\\&&-\frac{A_0(m_{\pi }^2) \bigg[8 \sqrt{2} c_{\theta } \left(4 m_K^2+m_{\pi }^2\right) s_{\theta }+40 m_{\pi }^2 c_{\theta }^2+s_{\theta }^2 \left(-64 m_K^2+18 m_{\eta
   '}^2+6 m_{\pi }^2+15 s\right)\bigg]}{240 F_{\pi }^4}\nonumber\\&&+\frac{A_0(m_{\eta }^2)}{720 F_{\pi }^4} \bigg[3 c_{\theta }^2 s_{\theta }^2 \left(128 m_K^2-18 m_{\eta '}^2+12 m_{\eta }^2-42 m_{\pi
   }^2-15 s\right)+8 \sqrt{2} c_{\theta }^3 \left(7 m_{\pi }^2-12 m_K^2\right) s_{\theta }\nonumber\\&&+4 \sqrt{2} c_{\theta } \left(m_{\pi }^2-6 m_K^2\right) s_{\theta }^3+40 m_{\pi }^2 c_{\theta
   }^4+40 \left(m_{\pi }^2-3 m_K^2\right) s_{\theta }^4\bigg]\nonumber\\&&+\frac{A_0(m_{\eta '}^2)}{720 F_{\pi }^4} \bigg[-160 c_{\theta }^4 m_K^2+160 \sqrt{2} c_{\theta }^3 \left(2
   m_K^2-m_{\pi }^2\right) s_{\theta }-120 c_{\theta }^2 \left(5 m_K^2-2 m_{\pi }^2\right) s_{\theta }^2\nonumber\\&&+4 \sqrt{2} c_{\theta } \left(62 m_K^2-27 m_{\pi }^2\right) s_{\theta }^3+s_{\theta
   }^4 \left(64 m_K^2-18 m_{\eta '}^2+34 m_{\pi }^2-45 s\right)\bigg]\,, \nonumber
\end{eqnarray}
\begin{eqnarray}
&&T_{00,TAD}^{\eta\eta\to\eta\eta}(s)= \frac{A_0(m_{\eta '}^2)}{108 F_{\pi }^4} \bigg[2 c_{\theta }^6 \left(16 m_K^2-7 m_{\pi }^2\right)+3 c_{\theta }^4 \left(19 m_{\pi }^2-32 m_K^2\right) s_{\theta }^2-4 \sqrt{2} c_{\theta }^3
   \left(8 m_K^2-5 m_{\pi }^2\right) s_{\theta }^3\nonumber\\&&+24 c_{\theta }^2 \left(3 m_K^2-2 m_{\pi }^2\right) s_{\theta }^4+24 \sqrt{2} c_{\theta } \left(2 m_K^2-m_{\pi }^2\right) s_{\theta }^5-6
   \sqrt{2} m_{\pi }^2 c_{\theta }^5 s_{\theta }+4 \left(4 m_K^2-m_{\pi }^2\right) s_{\theta }^6\bigg]\nonumber\\&&-\frac{A_0(m_K^2)}{45 F_{\pi }^4} \bigg\{5 c_{\theta }^6 \left(16 m_K^2-7
   m_{\pi }^2\right)+c_{\theta }^4 \big[8 m_K^2 \left(30 s_{\theta }^2+1\right)+9 m_{\eta }^2+4 m_{\pi }^2 \left(2-15 s_{\theta }^2\right)\big]\nonumber\\&&+20 \sqrt{2} c_{\theta }^5 \left(8 m_K^2-5
   m_{\pi }^2\right) s_{\theta }+\sqrt{2} c_{\theta }^3 s_{\theta } \big[m_K^2 \left(80 s_{\theta }^2-46\right)+m_{\pi }^2 \left(21-80 s_{\theta }^2\right)\big]\nonumber\\&&+10 c_{\theta }^2
   s_{\theta }^2 \big[m_K^2 \left(2 s_{\theta }^2-9\right)+m_{\pi }^2 \left(s_{\theta }^2+3\right)\big]+20 \sqrt{2} c_{\theta } \left(m_{\pi }^2-2 m_K^2\right) s_{\theta }^3-20 m_K^2
   s_{\theta }^4\bigg\}\nonumber\\&&+\frac{A_0(m_{\eta }^2)}{108 F_{\pi
   }^4} \bigg[c_{\theta }^6 \left(64 m_K^2-31 m_{\pi }^2\right)+6 \sqrt{2} c_{\theta }^5 \left(32 m_K^2-17 m_{\pi
   }^2\right) s_{\theta }+30 c_{\theta }^4 \left(16 m_K^2-7 m_{\pi }^2\right) s_{\theta }^2\nonumber\\&&+40 \sqrt{2} c_{\theta }^3 \left(8 m_K^2-5 m_{\pi }^2\right) s_{\theta }^3+60 c_{\theta }^2
   \left(4 m_K^2-m_{\pi }^2\right) s_{\theta }^4+48 \sqrt{2} c_{\theta } \left(m_K^2-m_{\pi }^2\right) s_{\theta }^5\nonumber\\&&+4 \left(2 m_K^2+m_{\pi }^2\right) s_{\theta }^6\bigg] +\frac{m_{\pi }^2 A_0(m_{\pi }^2) \left(c_{\theta }^4-4 \sqrt{2} c_{\theta }^3 s_{\theta }+12 c_{\theta }^2 s_{\theta }^2-8 \sqrt{2} c_{\theta } s_{\theta }^3+4 s_{\theta
   }^4\right)}{12 F_{\pi }^4}\,, \nonumber
   \end{eqnarray}
\begin{eqnarray}
&&T_{00,TAD}^{\eta\eta\to\eta\eta'}(s) = \frac{A_0(m_K^2)}{90 \sqrt{2} F_{\pi }^4} \bigg\{-c_{\theta }^3 s_{\theta } \left(260 m_K^2 s_{\theta }^2+212 m_K^2+9 m_{\eta '}^2+27 m_{\eta
   }^2-125 m_{\pi }^2 s_{\theta }^2-28 m_{\pi }^2\right)\nonumber\\&&+15 c_{\theta }^5 \left(8 m_K^2+m_{\pi }^2\right) s_{\theta }+\sqrt{2} c_{\theta }^4 \big[m_{\pi }^2 \left(20 s_{\theta
   }^2+21\right)-2 m_K^2 \left(70 s_{\theta }^2+23\right)\big]\nonumber\\&&-3 \sqrt{2} c_{\theta }^2 s_{\theta }^2 \big[m_K^2 \left(40 s_{\theta }^2-6\right)+m_{\pi }^2 \left(1-25 s_{\theta
   }^2\right)\big]-20 c_{\theta } s_{\theta }^3 \big[5 m_K^2 \left(s_{\theta }^2-1\right)+m_{\pi }^2 \left(3-2 s_{\theta }^2\right)\big]\nonumber\\&&-20 \sqrt{2} s_{\theta }^4 \big[m_K^2
   \left(s_{\theta }^2-2\right)-m_{\pi }^2 \left(s_{\theta }^2-1\right)\big]+15 \sqrt{2} c_{\theta }^6 \left(8 m_K^2-5 m_{\pi }^2\right)\bigg\}-\nonumber\\&&\frac{\left(8 m_K^2-5 m_{\pi }^2\right) \left(2 \sqrt{2} c_{\theta }^6-6
   c_{\theta }^5 s_{\theta }-3 \sqrt{2} c_{\theta }^4 s_{\theta }^2+11 c_{\theta }^3 s_{\theta }^3+3 \sqrt{2} c_{\theta }^2 s_{\theta }^4-6 c_{\theta } s_{\theta }^5-2 \sqrt{2} s_{\theta
   }^6\right) A_0(m_{\eta '}^2)}{54 \sqrt{2} F_{\pi }^4}\nonumber\\&&-\frac{A_0(m_{\eta }^2)}{54
   \sqrt{2} F_{\pi }^4} \bigg[\sqrt{2} c_{\theta }^6 \left(32 m_K^2-17 m_{\pi }^2\right)+c_{\theta }^5 \left(96
   m_K^2 s_{\theta }-39 m_{\pi }^2 s_{\theta }\right)-20 c_{\theta }^3 \left(8 m_K^2-5 m_{\pi }^2\right) s_{\theta }^3\nonumber\\&&-60 \sqrt{2} c_{\theta }^2 \left(2 m_K^2-m_{\pi }^2\right) s_{\theta
   }^4+24 c_{\theta } \left(m_{\pi }^2-3 m_K^2\right) s_{\theta }^5-15 \sqrt{2} m_{\pi }^2 c_{\theta }^4 s_{\theta }^2+8 \sqrt{2} \left(m_{\pi }^2-m_K^2\right) s_{\theta }^6\bigg]\nonumber\\&&+\frac{m_{\pi }^2 A_0(m_{\pi }^2) \left(\sqrt{2} c_{\theta }^4-5 c_{\theta }^3 s_{\theta }+3 \sqrt{2} c_{\theta }^2 s_{\theta }^2+2 c_{\theta } s_{\theta
   }^3-2 \sqrt{2} s_{\theta }^4\right)}{6 \sqrt{2} F_{\pi }^4} \,, \nonumber
   \end{eqnarray}
\begin{eqnarray}
&&T_{00,TAD}^{\eta\eta\to\eta'\eta'}(s) = -\frac{A_0(m_K^2)}{90
   F_{\pi }^4} \bigg\{10 c_{\theta }^6 \left(4 m_K^2-m_{\pi }^2\right)+c_{\theta }^2 s_{\theta }^2 \big[-\left(4 m_K^2-m_{\pi }^2\right) \left(5 s_{\theta }^2-24\right)+9
   m_{\eta '}^2+9 m_{\eta }^2\big]\nonumber\\&&+10 \sqrt{2} c_{\theta }^5 \left(m_{\pi }^2-4 m_K^2\right) s_{\theta }-5 c_{\theta }^4 \big[m_K^2 \left(4 s_{\theta }^2+6\right)-m_{\pi }^2
   \left(s_{\theta }^2+2\right)\big]+\sqrt{2} c_{\theta }^3 \left(6 m_K^2-m_{\pi }^2\right) s_{\theta }\nonumber\\&&+\sqrt{2} c_{\theta } s_{\theta }^3 \big[m_K^2 \left(40 s_{\theta
   }^2-6\right)+m_{\pi }^2 \left(1-10 s_{\theta }^2\right)\big]+10 s_{\theta }^4 \big[m_K^2 \left(4 s_{\theta }^2-3\right)-m_{\pi }^2 \left(s_{\theta }^2-1\right)\big]\bigg\}\nonumber\\&&+\frac{A_0(m_{\eta '}^2)}{108 F_{\pi }^4} \bigg[4 c_{\theta }^6 \left(4 m_K^2-m_{\pi }^2\right)+24 \sqrt{2} c_{\theta }^5 \left(m_{\pi }^2-2 m_K^2\right) s_{\theta }+24 c_{\theta
   }^4 \left(3 m_K^2-2 m_{\pi }^2\right) s_{\theta }^2\nonumber\\&&+4 \sqrt{2} c_{\theta }^3 \left(8 m_K^2-5 m_{\pi }^2\right) s_{\theta }^3+3 c_{\theta }^2 \left(19 m_{\pi }^2-32 m_K^2\right) s_{\theta
   }^4+6 \sqrt{2} m_{\pi }^2 c_{\theta } s_{\theta }^5+2 \left(16 m_K^2-7 m_{\pi }^2\right) s_{\theta }^6\bigg]\nonumber\\&&+\frac{A_0(m_{\eta }^2)}{108 F_{\pi }^4} \bigg[2 c_{\theta }^6
   \left(16 m_K^2-7 m_{\pi }^2\right)+3 c_{\theta }^4 \left(19 m_{\pi }^2-32 m_K^2\right) s_{\theta }^2-4 \sqrt{2} c_{\theta }^3 \left(8 m_K^2-5 m_{\pi }^2\right) s_{\theta }^3\nonumber\\&&+24 c_{\theta
   }^2 \left(3 m_K^2-2 m_{\pi }^2\right) s_{\theta }^4+24 \sqrt{2} c_{\theta } \left(2 m_K^2-m_{\pi }^2\right) s_{\theta }^5-6 \sqrt{2} m_{\pi }^2 c_{\theta }^5 s_{\theta }+4 \left(4
   m_K^2-m_{\pi }^2\right) s_{\theta }^6\bigg]\nonumber\\&&+\frac{m_{\pi }^2 A_0(m_{\pi }^2) \left(2 c_{\theta }^4-2 \sqrt{2} c_{\theta }^3 s_{\theta }-3 c_{\theta }^2
   s_{\theta }^2+2 \sqrt{2} c_{\theta } s_{\theta }^3+2 s_{\theta }^4\right)}{12 F_{\pi }^4}\,, \nonumber
\end{eqnarray}
\begin{eqnarray}
&&T_{00,TAD}^{\eta\eta'\to\eta\eta'}(s) = -\frac{A_0(m_K^2)}{45
   F_{\pi }^4} \bigg\{10 c_{\theta }^6 \left(4 m_K^2-m_{\pi }^2\right)+c_{\theta }^2 s_{\theta }^2 \big[-\left(4 m_K^2-m_{\pi }^2\right) \left(5 s_{\theta }^2-24\right)+9
   m_{\eta '}^2+9 m_{\eta }^2\big]\nonumber\\&&+10 \sqrt{2} c_{\theta }^5 \left(m_{\pi }^2-4 m_K^2\right) s_{\theta }-5 c_{\theta }^4 \big[m_K^2 \left(4 s_{\theta }^2+6\right)-m_{\pi }^2
   \left(s_{\theta }^2+2\right)\big]+\sqrt{2} c_{\theta }^3 \left(6 m_K^2-m_{\pi }^2\right) s_{\theta }\nonumber\\&&+\sqrt{2} c_{\theta } s_{\theta }^3 \big[m_K^2 \left(40 s_{\theta
   }^2-6\right)+m_{\pi }^2 \left(1-10 s_{\theta }^2\right)\big]+10 s_{\theta }^4 \big[m_K^2 \left(4 s_{\theta }^2-3\right)-m_{\pi }^2 \left(s_{\theta }^2-1\right)\big]\bigg\}\nonumber\\&&+\frac{A_0(m_{\eta '}^2)}{54 F_{\pi }^4} \bigg[4 c_{\theta }^6 \left(4 m_K^2-m_{\pi }^2\right)+24 \sqrt{2} c_{\theta }^5 \left(m_{\pi }^2-2 m_K^2\right) s_{\theta }+24 c_{\theta
   }^4 \left(3 m_K^2-2 m_{\pi }^2\right) s_{\theta }^2\nonumber\\&&+4 \sqrt{2} c_{\theta }^3 \left(8 m_K^2-5 m_{\pi }^2\right) s_{\theta }^3+3 c_{\theta }^2 \left(19 m_{\pi }^2-32 m_K^2\right) s_{\theta
   }^4+6 \sqrt{2} m_{\pi }^2 c_{\theta } s_{\theta }^5+2 \left(16 m_K^2-7 m_{\pi }^2\right) s_{\theta }^6\bigg]\nonumber\\&&+\frac{A_0(m_{\eta }^2)}{54 F_{\pi }^4} \bigg[2 c_{\theta }^6
   \left(16 m_K^2-7 m_{\pi }^2\right)+3 c_{\theta }^4 \left(19 m_{\pi }^2-32 m_K^2\right) s_{\theta }^2-4 \sqrt{2} c_{\theta }^3 \left(8 m_K^2-5 m_{\pi }^2\right) s_{\theta }^3\nonumber\\&&+24 c_{\theta
   }^2 \left(3 m_K^2-2 m_{\pi }^2\right) s_{\theta }^4+24 \sqrt{2} c_{\theta } \left(2 m_K^2-m_{\pi }^2\right) s_{\theta }^5-6 \sqrt{2} m_{\pi }^2 c_{\theta }^5 s_{\theta }+4 \left(4
   m_K^2-m_{\pi }^2\right) s_{\theta }^6\bigg]\nonumber\\&&+\frac{m_{\pi }^2 A_0(m_{\pi }^2) \left(2 c_{\theta }^4-2 \sqrt{2} c_{\theta }^3 s_{\theta }-3 c_{\theta }^2
   s_{\theta }^2+2 \sqrt{2} c_{\theta } s_{\theta }^3+2 s_{\theta }^4\right)}{6 F_{\pi }^4} \,, \nonumber
   \end{eqnarray}
\begin{eqnarray}
&&T_{00,TAD}^{\eta\eta'\to\eta'\eta'}(s) = \frac{A_0(m_K^2)}{90 \sqrt{2} F_{\pi }^4} \bigg\{c_{\theta } s_{\theta }^3 \left(120 m_K^2 s_{\theta }^2-212 m_K^2-27 m_{\eta '}^2-9 m_{\eta }^2+15
   m_{\pi }^2 s_{\theta }^2+28 m_{\pi }^2\right)\nonumber\\&&-20 c_{\theta }^5 \left(5 m_K^2-2 m_{\pi }^2\right) s_{\theta }+5 \sqrt{2} c_{\theta }^4 \big[8 m_K^2 \left(3 s_{\theta }^2-1\right)+m_{\pi
   }^2 \left(4-15 s_{\theta }^2\right)\big]\nonumber\\&&-5 c_{\theta }^3 s_{\theta } \big[4 m_K^2 \left(13 s_{\theta }^2-5\right)+m_{\pi }^2 \left(12-25 s_{\theta }^2\right)\big]+\sqrt{2} c_{\theta
   }^2 s_{\theta }^2 \big[2 m_K^2 \left(70 s_{\theta }^2-9\right)+m_{\pi }^2 \left(3-20 s_{\theta }^2\right)\big]\nonumber\\&&+\sqrt{2} s_{\theta }^4 \big[m_K^2 \left(46-120 s_{\theta }^2\right)+3
   m_{\pi }^2 \left(25 s_{\theta }^2-7\right)\big]+20 \sqrt{2} c_{\theta }^6 \left(m_K^2-m_{\pi }^2\right)\bigg\}\nonumber\\&&-\frac{A_0(m_{\eta '}^2)}{54 \sqrt{2} F_{\pi }^4} \bigg[8 \sqrt{2} c_{\theta }^6 \left(m_K^2-m_{\pi }^2\right)+24
   c_{\theta }^5 \left(m_{\pi }^2-3 m_K^2\right) s_{\theta }+60 \sqrt{2} c_{\theta }^4 \left(2 m_K^2-m_{\pi }^2\right) s_{\theta }^2\nonumber\\&&-20 c_{\theta }^3 \left(8 m_K^2-5 m_{\pi }^2\right)
   s_{\theta }^3+3 c_{\theta } \left(32 m_K^2-13 m_{\pi }^2\right) s_{\theta }^5+15 \sqrt{2} m_{\pi }^2 c_{\theta }^2 s_{\theta }^4+\sqrt{2} \left(17 m_{\pi }^2-32 m_K^2\right) s_{\theta
   }^6\bigg]\nonumber\\&&-\frac{\left(8 m_K^2-5 m_{\pi }^2\right) A_0(m_{\eta }^2) \left(2 \sqrt{2} c_{\theta }^6-6 c_{\theta }^5 s_{\theta }-3 \sqrt{2} c_{\theta
   }^4 s_{\theta }^2+11 c_{\theta }^3 s_{\theta }^3+3 \sqrt{2} c_{\theta }^2 s_{\theta }^4-6 c_{\theta } s_{\theta }^5-2 \sqrt{2} s_{\theta }^6\right)}{54 \sqrt{2} F_{\pi }^4}\nonumber\\&&+\frac{m_{\pi
   }^2 A_0(m_{\pi }^2) \left(2 \sqrt{2} c_{\theta }^4+2 c_{\theta }^3 s_{\theta }-3 \sqrt{2} c_{\theta }^2 s_{\theta }^2-5 c_{\theta } s_{\theta }^3-\sqrt{2} s_{\theta
   }^4\right)}{6 \sqrt{2} F_{\pi }^4}\,, \nonumber
   \end{eqnarray}
\begin{eqnarray}
&&T_{00,TAD}^{\eta'\eta'\to\eta'\eta'}(s)= -\frac{A_0(m_K^2)}{45 F_{\pi }^4} \bigg\{10 c_{\theta }^4 \big[2 m_K^2 \left(s_{\theta }^2-1\right)+m_{\pi }^2 s_{\theta }^2\big]-20 \sqrt{2} c_{\theta }^3 s_{\theta } \big[m_K^2 \left(4
   s_{\theta }^2-2\right)+m_{\pi }^2 \left(1-4 s_{\theta }^2\right)\big]\nonumber\\&&+30 c_{\theta }^2 s_{\theta }^2 \big[m_K^2 \left(8 s_{\theta }^2-3\right)+m_{\pi }^2 \left(1-2 s_{\theta
   }^2\right)\big]+\sqrt{2} c_{\theta } s_{\theta }^3 \big[m_K^2 \left(46-160 s_{\theta }^2\right)+m_{\pi }^2 \left(100 s_{\theta }^2-21\right)\big]\nonumber\\&&+s_{\theta }^4 \big[m_K^2 \left(80
   s_{\theta }^2+8\right)+9 m_{\eta '}^2+m_{\pi }^2 \left(8-35 s_{\theta }^2\right)\big]\bigg\}\nonumber\\&&+\frac{A_0(m_{\eta '}^2)}{108 F_{\pi }^4} \bigg[4 c_{\theta }^6 \left(2
   m_K^2+m_{\pi }^2\right)-48 \sqrt{2} c_{\theta }^5 \left(m_K^2-m_{\pi }^2\right) s_{\theta }+60 c_{\theta }^4 \left(4 m_K^2-m_{\pi }^2\right) s_{\theta }^2\nonumber\\&&-40 \sqrt{2} c_{\theta }^3
   \left(8 m_K^2-5 m_{\pi }^2\right) s_{\theta }^3+30 c_{\theta }^2 \left(16 m_K^2-7 m_{\pi }^2\right) s_{\theta }^4-6 \sqrt{2} c_{\theta } \left(32 m_K^2-17 m_{\pi }^2\right) s_{\theta
   }^5\nonumber\\&&+\left(64 m_K^2-31 m_{\pi }^2\right) s_{\theta }^6\bigg]+\frac{m_{\pi }^2 A_0(m_{\pi }^2) \left(4 c_{\theta }^4+8 \sqrt{2} c_{\theta }^3 s_{\theta }+12 c_{\theta }^2 s_{\theta }^2+4 \sqrt{2} c_{\theta }
   s_{\theta }^3+s_{\theta }^4\right)}{12 F_{\pi }^4}\nonumber\\&&+\frac{A_0(m_{\eta }^2)}{108 F_{\pi }^4} \bigg[4 c_{\theta }^6 \left(4 m_K^2-m_{\pi }^2\right)+24 \sqrt{2}
   c_{\theta }^5 \left(m_{\pi }^2-2 m_K^2\right) s_{\theta }+24 c_{\theta }^4 \left(3 m_K^2-2 m_{\pi }^2\right) s_{\theta }^2\nonumber\\&&+4 \sqrt{2} c_{\theta }^3 \left(8 m_K^2-5 m_{\pi }^2\right)
   s_{\theta }^3+3 c_{\theta }^2 \left(19 m_{\pi }^2-32 m_K^2\right) s_{\theta }^4+6 \sqrt{2} m_{\pi }^2 c_{\theta } s_{\theta }^5+2 \left(16 m_K^2-7 m_{\pi }^2\right) s_{\theta
   }^6\bigg]\,.
\end{eqnarray}

For the $(I,J)=(1,0)$ case, the expressions from the tadpole diagrams take the form
\begin{eqnarray}\label{eq.pw10tad}
&&T_{10,TAD}^{\pi\eta\to\pi\eta}(s) = \frac{A_0(m_K^2)}{45 F_{\pi }^4} \bigg[-c_{\theta }^2 \left(16 m_K^2+3 m_{\eta }^2+30 m_{\pi }^2  s_{\theta }^2+6m_{\pi}^2\right)+\sqrt{2} c_{\theta } \left(2 m_K^2+3 m_{\pi }^2\right)
   s_{\theta }\nonumber\\&&-15 m_{\pi }^2 c_{\theta }^4+30 \sqrt{2} m_{\pi }^2 c_{\theta }^3 s_{\theta }+10 m_K^2 s_{\theta }^2\bigg]+\frac{m_{\pi }^2 \left(2 c_{\theta }^4-2 \sqrt{2}
   c_{\theta }^3 s_{\theta }-3 c_{\theta }^2 s_{\theta }^2+2 \sqrt{2} c_{\theta } s_{\theta }^3+2 s_{\theta }^4\right) A_0(m_{\eta '}^2)}{18 F_{\pi }^4}\nonumber\\&&+\frac{m_{\pi }^2
   A_0(m_{\eta }^2) \left(c_{\theta }^4-4 \sqrt{2} c_{\theta }^3 s_{\theta }+12 c_{\theta }^2 s_{\theta }^2-8 \sqrt{2} c_{\theta } s_{\theta }^3+4 s_{\theta }^4\right)}{18 F_{\pi
   }^4}+\frac{m_{\pi }^2 A_0(m_{\pi }^2) \left(c_{\theta }^2-2 \sqrt{2} c_{\theta } s_{\theta }+2 s_{\theta }^2\right)}{18 F_{\pi }^4}\,,\nonumber
   \end{eqnarray}
\begin{eqnarray}
&&T_{10,TAD}^{\pi\eta\to K\bar{K}}(s) = -\frac{A_0(m_{\eta '}^2)}{360 \sqrt{6} F_{\pi
   }^4} \bigg[-40 c_{\theta }^3 \left(m_K^2-m_{\pi }^2\right)+3 c_{\theta } s_{\theta }^2 \left(-16 m_K^2-12 m_{\eta '}^2+9 m_{\eta }^2-21 m_{\pi }^2+15
   s\right)\nonumber\\&&-12 \sqrt{2} c_{\theta }^2 \left(2 m_K^2+3 m_{\pi }^2\right) s_{\theta }+4 \sqrt{2} \left(8 m_K^2+7 m_{\pi }^2\right) s_{\theta }^3\bigg]\nonumber\\&&-\frac{A_0(m_K^2)}{180 \sqrt{6} F_{\pi }^4} \bigg[15 c_{\theta }^3 \left(8 m_K^2+3 m_{\eta }^2+m_{\pi }^2-9 s\right)-2 c_{\theta } \left(16 m_K^2+3 m_{\eta }^2+41 m_{\pi }^2-30 s\right)\nonumber\\&&+30 \sqrt{2}
   c_{\theta }^2 \left(2 m_K^2+m_{\pi }^2\right) s_{\theta }+16 \sqrt{2} \left(m_{\pi }^2-m_K^2\right) s_{\theta }\bigg]\nonumber\\&&-\frac{A_0(m_{\pi }^2)
   \big[c_{\theta } \left(32 m_K^2-9 m_{\eta }^2+25 m_{\pi }^2-15 s\right)+4 \sqrt{2} \left(4 m_K^2-m_{\pi }^2\right) s_{\theta }\big]}{72 \sqrt{6} F_{\pi }^4}\nonumber\\&&-\frac{A_0\left(m_{\eta
   }^2\right)}{360 \sqrt{6} F_{\pi }^4} \bigg[c_{\theta }^3 \left(-128 m_K^2-9 m_{\eta }^2+17 m_{\pi }^2+45 s\right)-24 \sqrt{2} c_{\theta }^2 \left(m_K^2-m_{\pi }^2\right) s_{\theta }\nonumber\\&&-120 c_{\theta }
   \left(m_K^2-m_{\pi }^2\right) s_{\theta }^2-40 \sqrt{2} \left(2 m_K^2+m_{\pi }^2\right) s_{\theta }^3\bigg]\,,\nonumber
   \end{eqnarray}
\begin{eqnarray}
&&T_{10,TAD}^{\pi\eta\to\pi\eta'}(s)= \frac{A_0(m_K^2)}{90 F_{\pi }^4} \bigg[-\sqrt{2} c_{\theta }^2 \left(2 m_K^2+3 m_{\pi }^2\right)-c_{\theta } s_{\theta } \left(52 m_K^2+3 m_{\eta '}^2+3 m_{\eta }^2-15 m_{\pi }^2 s_{\theta
   }^2+12 m_{\pi }^2\right)\nonumber\\&&-15 \sqrt{2} m_{\pi }^2 c_{\theta }^4+15 m_{\pi }^2 c_{\theta }^3 s_{\theta }+\sqrt{2} s_{\theta }^2 \left(2 m_K^2+15 m_{\pi }^2  s_{\theta
   }^2+3m_{\pi}^2\right)\bigg] +\frac{m_{\pi }^2 A_0(m_{\pi }^2) \left(\sqrt{2} c_{\theta
   }^2-c_{\theta } s_{\theta }-\sqrt{2} s_{\theta }^2\right)}{18 F_{\pi }^4} \nonumber\\&&+\frac{m_{\pi }^2 \left(2 \sqrt{2} c_{\theta }^4+2 c_{\theta }^3 s_{\theta }-3 \sqrt{2} c_{\theta }^2 s_{\theta }^2-5 c_{\theta } s_{\theta
   }^3-\sqrt{2} s_{\theta }^4\right) A_0(m_{\eta '}^2)}{18 F_{\pi }^4}\nonumber\\&&+\frac{m_{\pi }^2 A_0(m_{\eta }^2) \left(\sqrt{2} c_{\theta }^4-5 c_{\theta }^3 s_{\theta }+3
   \sqrt{2} c_{\theta }^2 s_{\theta }^2+2 c_{\theta } s_{\theta }^3-2 \sqrt{2} s_{\theta }^4\right)}{18 F_{\pi }^4}\,,\nonumber
   \end{eqnarray}
\begin{eqnarray}
&&T_{10,TAD}^{K\bar{K}\to K\bar{K}}(s)=\frac{A_0(m_{\eta '}^2) \bigg[20 c_{\theta }^2 m_K^2+2 \sqrt{2} c_{\theta } \left(3 m_{\pi }^2-8 m_K^2\right) s_{\theta }-s_{\theta }^2 \left(8 m_K^2+24 m_{\eta '}^2+3 m_{\pi
   }^2+15 s\right)\bigg]}{180 F_{\pi }^4}\nonumber\\&&-\frac{A_0(m_{\eta }^2) \bigg[c_{\theta }^2 \left(8 m_K^2+24 m_{\eta }^2+3 m_{\pi }^2+15 s\right)+2 \sqrt{2} c_{\theta } \left(3 m_{\pi
   }^2-8 m_K^2\right) s_{\theta }-20 m_K^2 s_{\theta }^2\bigg]}{180 F_{\pi }^4}\nonumber\\&&-\frac{A_0(m_K^2) \left(20 m_K^2-6 s\right)}{36 F_{\pi }^4}-\frac{A_0(m_{\pi }^2)
   \left(-4 m_K^2+m_{\pi }^2+6 s\right)}{36 F_{\pi }^4}\,,\nonumber \\
\end{eqnarray}
\begin{eqnarray}
&&T_{10,TAD}^{K\bar{K}\to \pi\eta'}(s)= -\frac{A_0(m_{\pi }^2) \bigg[4 \sqrt{2} c_{\theta } \left(m_{\pi }^2-4 m_K^2\right)+s_{\theta } \left(32 m_K^2-9 m_{\eta '}^2+25 m_{\pi }^2-15 s\right)\bigg]}{72 \sqrt{6} F_{\pi
   }^4}\nonumber\\&&-\frac{A_0(m_{\eta }^2)}{360 \sqrt{6} F_{\pi
   }^4} \bigg[-4 \sqrt{2} c_{\theta }^3 \left(8 m_K^2+7 m_{\pi }^2\right)-3 c_{\theta }^2 s_{\theta } \left(16 m_K^2-9 m_{\eta '}^2+12 m_{\eta }^2+21
   m_{\pi }^2-15 s\right)\nonumber\\&&+12 \sqrt{2} c_{\theta } \left(2 m_K^2+3 m_{\pi }^2\right) s_{\theta }^2+40 \left(m_{\pi }^2-m_K^2\right) s_{\theta }^3\bigg]\nonumber\\&&-\frac{A_0(m_{\eta '}^2)}{360 \sqrt{6} F_{\pi }^4} \bigg[40 \sqrt{2} c_{\theta }^3 \left(2 m_K^2+m_{\pi }^2\right)-120 c_{\theta }^2 \left(m_K^2-m_{\pi }^2\right) s_{\theta }\nonumber\\&&+24 \sqrt{2} c_{\theta
   } \left(m_K^2-m_{\pi }^2\right) s_{\theta }^2+s_{\theta }^3 \left(-128 m_K^2-9 m_{\eta '}^2+17 m_{\pi }^2+45 s\right)\bigg]\nonumber\\&&-\frac{A_0(m_K^2)}{180 \sqrt{6} F_{\pi }^4}
   \bigg\{s_{\theta } \Big[8 m_K^2 \left(15 s_{\theta }^2-4\right)+\left(45 s_{\theta }^2-6\right) m_{\eta '}^2+15 m_{\pi }^2 s_{\theta }^2-82 m_{\pi }^2+60 s-135 \text{ss}_{\theta
   }^2\Big]\nonumber\\&&-2 \sqrt{2} c_{\theta } \left(30m_K^2 s_{\theta }^2-8m_K^2+15m_{\pi }^2  s_{\theta }^2+8m_{\pi}^2\right)\bigg\}\,,\nonumber
   \end{eqnarray}
\begin{eqnarray}
&&T_{10,TAD}^{\pi\eta'\to\pi\eta'}(s)= \frac{A_0(m_K^2)}{45 F_{\pi }^4} \bigg\{10 c_{\theta }^2 \left(m_K^2-3 m_{\pi }^2 s_{\theta }^2\right)-\sqrt{2} c_{\theta } s_{\theta } \big[2 m_K^2+3 m_{\pi }^2 \left(10 s_{\theta
   }^2+1\right)\big]\nonumber\\&&-s_{\theta }^2 \big[16 m_K^2+3 m_{\eta '}^2+3 m_{\pi }^2 \left(5 s_{\theta }^2+2\right)\big]\bigg\}+\frac{m_{\pi }^2
   A_0(m_{\eta }^2) \left(2 c_{\theta }^4-2 \sqrt{2} c_{\theta }^3 s_{\theta }-3 c_{\theta }^2 s_{\theta }^2+2 \sqrt{2} c_{\theta } s_{\theta }^3+2 s_{\theta }^4\right)}{18
   F_{\pi }^4}\nonumber\\&&+\frac{m_{\pi }^2 \left(4 c_{\theta }^4+8
   \sqrt{2} c_{\theta }^3 s_{\theta }+12 c_{\theta }^2 s_{\theta }^2+4 \sqrt{2} c_{\theta } s_{\theta }^3+s_{\theta }^4\right) A_0(m_{\eta '}^2)}{18 F_{\pi }^4}+\frac{m_{\pi }^2 A_0(m_{\pi }^2) \left(2 c_{\theta }^2+2 \sqrt{2} c_{\theta } s_{\theta }+s_{\theta }^2\right)}{18 F_{\pi }^4}\,. \nonumber \\
\end{eqnarray}

For the $(I,J)=(1/2,0)$ case, the expressions from the tadpole diagrams read
\begin{eqnarray}\label{eq.pw1d20tad}
&&T_{\frac{1}{2}0,TAD}^{K\pi\to K\pi}(s) = \frac{A_0(m_{\eta '}^2)}{1440 F_{\pi }^4 s} \bigg\{80 s c_{\theta }^2 \left(m_K^2+m_{\pi }^2\right)-8 \sqrt{2} s c_{\theta } \left(2 m_K^2-7 m_{\pi }^2\right) s_{\theta }\nonumber\\&&+s_{\theta }^2 \big[-2 m_K^2
   \left(15 m_{\pi }^2+19 s\right)+15 m_K^4-3 s \left(8 m_{\eta '}^2+35 s\right)+82 m_{\pi }^2 s+15 m_{\pi }^4\big]\bigg\}\nonumber\\&&+\frac{A_0(m_{\eta }^2)}{1440 F_{\pi }^4 s}
   \bigg\{c_{\theta }^2 \big[-2 m_K^2 \left(15 m_{\pi }^2+19 s\right)+15 m_K^4-3 s \left(8 m_{\eta }^2+35 s\right)+82 m_{\pi }^2 s+15 m_{\pi }^4\big]\nonumber\\&&+8 \sqrt{2} s c_{\theta } \left(2
   m_K^2-7 m_{\pi }^2\right) s_{\theta }+80 \left(m_K^2+m_{\pi }^2\right) \text{ss}_{\theta }^2\bigg\}\nonumber\\&&+\frac{A_0(m_K^2) \big[-2 m_K^2 \left(7 m_{\pi }^2+11
   s\right)+7 m_K^4+2 m_{\pi }^2 s+7 m_{\pi }^4-17 s^2\big]}{144 F_{\pi }^4 s}\nonumber\\&&+\frac{A_0(m_{\pi }^2) \big[-2 m_K^2 \left(19 m_{\pi }^2+7 s\right)+19 m_K^4-94 m_{\pi }^2 s+19
   m_{\pi }^4-5 s^2\big]}{288 F_{\pi }^4 s}\,,\nonumber \\
   \end{eqnarray}
\begin{eqnarray}
&&T_{\frac{1}{2}0,TAD}^{K\pi\to K\eta}(s) =\frac{A_0(m_{\eta '}^2)}{1440 F_{\pi }^4} \bigg\{80 c_{\theta }^3 \left(m_K^2-m_{\pi }^2\right)+24 \sqrt{2} c_{\theta }^2 \left(2 m_K^2+3 m_{\pi }^2\right) s_{\theta
   }-8 \sqrt{2} \left(8 m_K^2+7 m_{\pi }^2\right) s_{\theta }^3\nonumber\\&&+3 c_{\theta } s_{\theta }^2 \bigg[15 \frac{\left(m_K^2-m_{\pi }^2+s\right)\left(m_K^2-m_{\eta }^2+s\right)}{s}
   -28 m_K^2+24 m_{\eta '}^2-18 m_{\eta }^2+42 m_{\pi }^2\bigg]\bigg\}\nonumber\\&&-\frac{A_0(m_K^2)}{720 F_{\pi }^4} \bigg\{15 c_{\theta }^3 \bigg[9 \frac{\left(m_K^2-m_{\pi
   }^2+s\right)\left(m_K^2-m_{\eta }^2+s\right)}{s} -20 m_K^2+6 m_{\eta }^2+2 m_{\pi }^2\bigg]+32 \sqrt{2} \left(m_{\pi }^2-m_K^2\right) s_{\theta }\nonumber\\&&-4 c_{\theta } \bigg[15  \frac{\left(m_K^2-m_{\pi
   }^2+s\right)\left(m_K^2-m_{\eta }^2+s\right)}{s}-44 m_K^2+3 m_{\eta }^2+41 m_{\pi }^2\bigg]+60 \sqrt{2} c_{\theta }^2 \left(2 m_K^2+m_{\pi }^2\right) s_{\theta} \bigg\}
   \nonumber\\&&
   -\frac{A_0(m_{\pi }^2)}{288
   F_{\pi }^4} \Bigg\{c_{\theta } \bigg[15  \frac{\left(m_K^2-m_{\pi
   }^2+s\right)\left(m_K^2-m_{\eta }^2+s\right)}{s} +4 m_K^2-18 m_{\eta }^2+50 m_{\pi }^2\bigg] +8 \sqrt{2} \left(4 m_K^2-m_{\pi }^2\right) s_{\theta }\Bigg\}\nonumber\\&&+\frac{A_0(m_{\eta }^2)}{1440 F_{\pi }^4} \Bigg\{c_{\theta }^3 \bigg[45  \frac{\left(m_K^2-m_{\pi }^2+s\right)\left(m_K^2-m_{\eta }^2+s\right)}{s}  +76
   m_K^2+18 m_{\eta }^2-34 m_{\pi }^2\bigg]\nonumber\\&&+48 \sqrt{2} c_{\theta }^2 \left(m_K^2-m_{\pi }^2\right) s_{\theta }+240 c_{\theta } \left(m_K^2-m_{\pi }^2\right) s_{\theta }^2+80 \sqrt{2}
   \left(2 m_K^2+m_{\pi }^2\right) s_{\theta }^3\Bigg\} \,,\nonumber
\end{eqnarray}
\begin{eqnarray}
&&T_{\frac{1}{2}0,TAD}^{K\pi\to K\eta'}(s)= \frac{A_0(m_K^2)}{720 F_{\pi }^4} \bigg\{4 \sqrt{2} c_{\theta } \big[m_K^2 \left(30 s_{\theta }^2-8\right)+m_{\pi }^2 \left(15 s_{\theta }^2+8\right)\big]\nonumber\\&&+
    15s_{\theta }(4-9s_{\theta }^2)\frac{\left(m_K^2-m_{\pi }^2+s\right)\left(m_K^2-m_{\eta '}^2+s\right)}{s}   \nonumber\\&&+s_{\theta } \left[4 m_K^2 \left(75 s_{\theta }^2-44\right)+\left(12-90 s_{\theta }^2\right) m_{\eta '}^2-30 m_{\pi }^2 s_{\theta }^2+164 m_{\pi
   }^2\right]\bigg\}\nonumber\\&&-\frac{A_0(m_{\pi }^2)}{288 F_{\pi }^4} \bigg[8 \sqrt{2} c_{\theta } \left(m_{\pi }^2-4 m_K^2\right)+15 s_{\theta } \frac{\left(m_K^2-m_{\pi
   }^2+s\right)\left(m_K^2-m_{\eta '}^2+s\right)}{s} +s_{\theta } \left(4 m_K^2-18 m_{\eta '}^2+50 m_{\pi }^2\right)\bigg]\nonumber\\&&+\frac{A_0(m_{\eta }^2)}{1440 F_{\pi }^4}
   \bigg\{8 \sqrt{2} c_{\theta }^3 \left(8 m_K^2+7 m_{\pi }^2\right)-24 \sqrt{2} c_{\theta } \left(2 m_K^2+3 m_{\pi }^2\right) s_{\theta }^2+80 \left(m_K^2-m_{\pi
   }^2\right) s_{\theta }^3\nonumber\\&&+3 c_{\theta }^2 s_{\theta } \bigg[15 \frac{\left(m_K^2-m_{\pi }^2+s\right)\left(m_K^2-m_{\eta
   '}^2+s\right)}{s} -28 m_K^2-18 m_{\eta '}^2+24 m_{\eta }^2+42 m_{\pi }^2\bigg]\bigg\}\nonumber\\&&-\frac{A_0(m_{\eta '}^2)}{1440 F_{\pi }^4} \bigg\{80 \sqrt{2} c_{\theta }^3 \left(2 m_K^2+m_{\pi }^2\right)-240 c_{\theta }^2 \left(m_K^2-m_{\pi
   }^2\right) s_{\theta }+48 \sqrt{2} c_{\theta } \left(m_K^2-m_{\pi }^2\right) s_{\theta }^2\nonumber\\&&-s_{\theta }^3 \bigg[45  \frac{\left(m_K^2-m_{\pi }^2+s\right)\left(m_K^2-m_{\eta '}^2+s\right)}{s}
    +76 m_K^2+18 m_{\eta '}^2-34 m_{\pi }^2\bigg]\bigg\} \,,\nonumber
   \end{eqnarray}
\begin{eqnarray}
&&T_{\frac{1}{2}0,TAD}^{K\eta\to K\eta}(s)= -\frac{A_0(m_{\eta }^2)}{1440 F_{\pi }^4 s} \bigg[c_{\theta }^4 \left(-90 m_{\eta }^2  m_K^2-126 sm_{\eta }^2+38 s m_K^2+45 m_K^4+45 m_{\eta }^4+68 m_{\pi }^2 s+45 s^2\right)\nonumber\\&&-8 \sqrt{2} s c_{\theta
   }^3 \left(62 m_K^2-27 m_{\pi }^2\right) s_{\theta }-320 \sqrt{2} s c_{\theta } \left(2 m_K^2-m_{\pi }^2\right) s_{\theta }^3 -240 c_{\theta }^2 \left(5 m_K^2-2 m_{\pi }^2\right)
   s\,s_{\theta }^2-320 s \, m_K^2 s_{\theta }^4\bigg]\nonumber\\&&+\frac{A_0(m_{\eta '}^2)}{1440 F_{\pi }^4
   s} \bigg\{-3 c_{\theta }^2 s_{\theta }^2 \big[-6 m_{\eta }^2 \left(5 m_K^2+11
   s\right)+226 s m_K^2+15 m_K^4\big]-16 \sqrt{2} s c_{\theta } \left(12 m_K^2-7 m_{\pi }^2\right) s_{\theta }^3 \nonumber\\&&-3 c_{\theta }^2 s_{\theta }^2 \big[+15 m_{\eta }^4+3 s \left(8 m_{\eta '}^2-28 m_{\pi }^2+5 s\right)\big]+80 s c_{\theta }^4 \left(3 m_K^2-m_{\pi }^2\right)+8 \sqrt{2} s c_{\theta }^3 \left(m_{\pi }^2-6 m_K^2\right) s_{\theta }-80 m_{\pi }^2 s s_{\theta }^4\bigg\}\nonumber\\&&+\frac{A_0(m_K^2)}{240 F_{\pi }^4
   s} \bigg\{10 c_{\theta }^4 \big[-6 m_{\eta }^2 \left(3 m_K^2+s\right)-18 s m_K^2+9 m_K^4+9 m_{\eta }^4+4s  m_{\pi }^2+9 s^2\big]\nonumber\\&&-c_{\theta }^2
   \big[m_{\eta }^2 \left(42 s-90 m_K^2\right)+2 s m_K^2 \left(80 s_{\theta }^2-13\right)+45 m_K^4+45 m_{\eta }^4+4s  m_{\pi }^2+45 s^2\big]\nonumber\\&&-80 \sqrt{2} s c_{\theta }^3
   \left(2 m_K^2-m_{\pi }^2\right) s_{\theta }+8 \sqrt{2} s c_{\theta } \left(6 m_K^2-m_{\pi }^2\right) s_{\theta }+80 m_K^2 s\,s_{\theta }^2\bigg\}\nonumber\\&&-\frac{A_0(m_{\pi }^2)}{480 F_{\pi }^4 s} \bigg\{c_{\theta }^2 \big[-6 m_{\eta }^2 \left(5 m_K^2+11 s\right)+98 s m_K^2+15 m_K^4+15 m_{\eta }^4+ 15 s^2-12s m_{\pi }^2\big]\nonumber \\ &&+16
   \sqrt{2} s c_{\theta } \left(4 m_K^2+m_{\pi }^2\right) s_{\theta }-80 m_{\pi }^2 s\,s_{\theta }^2\bigg\} \,,\nonumber \\
\end{eqnarray}
\begin{eqnarray}
&&T_{\frac{1}{2}0,TAD}^{K\eta\to K\eta'}(s) = \frac{ A_0(m_K^2)}{240 F_{\pi }^4}\Bigg\{-64 \sqrt{2} c_{\theta }^2 m_K^2+64 \sqrt{2} s_{\theta }^2 m_K^2+136 c_{\theta } s_{\theta } m_K^2+24 \sqrt{2} c_{\theta }^2 m_{\pi }^2-24 \sqrt{2} m_{\pi }^2 s_{\theta }^2\nonumber\\&&-24
   c_{\theta } m_{\pi }^2 s_{\theta }-96 c_{\theta } m_{\eta }^2 s_{\theta }-96 c_{\theta } m_{\eta '}^2 s_{\theta }-90 c_{\theta }s_{\theta }  \frac{\left(m_K^2-m_{\eta }^2+s\right)\left(m_K^2-m_{\eta '}^2+s\right)}{s}\nonumber\\&&+5 \left(c_{\theta }^2+s_{\theta }^2+1\right) \left(8 \sqrt{2}  m_K^2c_{\theta }^2-4\sqrt{2}m_{\pi }^2c_{\theta}^2 +4 \sqrt{2} m_{\pi }^2s_{\theta}^2-8\sqrt{2} m_K^2 s_{\theta }^2\right)+5 \left(c_{\theta }^2+s_{\theta }^2+1\right)\nonumber\\&& \bigg[-20
   m_K^2+4 m_{\pi }^2+6 m_{\eta }^2+6 m_{\eta '}^2+9  \frac{\left(m_K^2-m_{\eta }^2+s\right)\left(m_K^2-m_{\eta '}^2+s\right)}{s}  \bigg] s_{\theta }c_{\theta }\Bigg\}\nonumber\\&&+\frac{A_0(m_{\pi }^2)}{480 F_{\pi }^4}\bigg\{8 \sqrt{2} \left(4 m_K^2+m_{\pi }^2\right) c_{\theta
   }^2-8 \sqrt{2} \left(4 m_K^2+m_{\pi }^2\right) s_{\theta }^2\nonumber\\&&-\bigg[68 m_K^2+68 m_{\pi }^2-18 m_{\eta }^2-18 m_{\eta '}^2+15 \frac{\left(m_K^2-m_{\eta }^2+s\right)\left(m_K^2-m_{\eta '}^2+s\right)}{s} \bigg]
   s_{\theta } c_{\theta }\bigg\} \nonumber\\&&-\frac{ A_0(m_{\eta }^2)}{1440 F_{\pi }^4}\bigg[8 \sqrt{2} \left(8 m_K^2-3 m_{\pi
   }^2\right) c_{\theta }^4+48 \sqrt{2} \left(m_K^2-m_{\pi }^2\right) s_{\theta }^2 c_{\theta }^2-80 \left(5 m_K^2-3 m_{\pi }^2\right) s_{\theta }^3 c_{\theta
   }\nonumber\\&&+80 \sqrt{2} \left(m_{\pi }^2-2 m_K^2\right) s_{\theta }^4+s_{\theta } c_{\theta }^3\left(428 m_K^2-172 m_{\pi }^2+18 m_{\eta }^2-54 m_{\eta '}^2\right)\nonumber\\&&+ 45s_{\theta } c_{\theta }^3\frac{\left(m_K^2-m_{\eta }^2+s\right)\left(m_K^2-m_{\eta'}^2+s\right)}{s} \bigg]\nonumber\\&&+\frac{A_0(m_{\eta '}^2)}{1440 F_{\pi }^4}\bigg\{80 \sqrt{2} \left(m_{\pi }^2-2 m_K^2\right) c_{\theta }^4+80
   \left(5 m_K^2-3 m_{\pi }^2\right) s_{\theta } c_{\theta }^3\nonumber\\&&+48 \sqrt{2} \left(m_K^2-m_{\pi}^2\right) s_{\theta }^2 c_{\theta }^2 +8 \sqrt{2}\left(8 m_K^2-3 m_{\pi }^2\right) s_{\theta }^4\nonumber\\&&+s_{\theta }^3 c_{\theta }\bigg[-428 m_K^2+172 m_{\pi }^2+54
   m_{\eta }^2-18 m_{\eta '}^2-45\frac{\left(m_K^2-m_{\eta }^2+s\right)\left(m_K^2-m_{\eta '}^2+s\right)}{s}\bigg]  \bigg\}  \,,\nonumber
\end{eqnarray}
\begin{eqnarray}
&&T_{\frac{1}{2}0,TAD}^{K\eta'\to K\eta'}(s)=  \frac{A_0(m_{\eta '}^2)}{1440 F_{\pi }^4 s}\bigg[ 320 s c_{\theta }^4 m_K^2-320 \sqrt{2} s c_{\theta }^3 \left(2 m_K^2-m_{\pi }^2\right) s_{\theta }-8 \sqrt{2} s c_{\theta } \left(62 m_K^2-27 m_{\pi
   }^2\right) s_{\theta }^3\nonumber\\&&+240 c_{\theta }^2 \left(5 m_K^2-2 m_{\pi }^2\right) s\,s_{\theta }^2-s_{\theta }^4 \left(-90 m_K^2m_{\eta '}^2-126 s m_{\eta '}^2+38 s m_K^2+45 m_K^4+45
   m_{\eta '}^4+68 m_{\pi }^2 s+45 s^2\right)\bigg]\nonumber\\&&-\frac{A_0(m_K^2)}{240 F_{\pi }^4 s} \bigg\{80 s c_{\theta }^2 m_K^2 \left(2 s_{\theta }^2-1\right)-8 \sqrt{2} s c_{\theta }
   s_{\theta } \left(20m_K^2 s_{\theta }^2-6m_K^2+m_{\pi }^2 -10m_{\pi}^2 s_{\theta }^2\right)\nonumber\\&&+s_{\theta }^2 \big[ 6 m_{\eta '}^2 \left( 30  m_K^2 s_{\theta
   }^2-15m_K^2 +10 s\, s_{\theta }^2+7s \right)+45m_K^4 \left(1-2s_{\theta }^2\right)\big] \nonumber\\&&+s_{\theta }^2\big[2 s m_K^2 \left(90 s_{\theta }^2-13\right)+\left(45-90 s_{\theta }^2\right) m_{\eta '}^4+s
   \left(4m_{\pi }^2-40m_{\pi}^2 s_{\theta }^2+45 s\, -90 s\, s_{\theta }^2\right)\big]\bigg\}\nonumber\\&&+\frac{A_0(m_{\pi }^2)}{480 F_{\pi }^4 s} \bigg[16 \sqrt{2} s
   c_{\theta } \left(4 m_K^2+m_{\pi }^2\right) s_{\theta }+80 m_{\pi }^2 s c_{\theta }^2\nonumber\\&&+s_{\theta }^2 \left(30 m_K^2m_{\eta '}^2+66 s m_{\eta '}^2-98 s m_K^2-15 m_K^4-15 m_{\eta '}^4+12 s
    m_{\pi }^2-15 s^2\right)\bigg]\nonumber\\&&-\frac{A_0(m_{\eta }^2)}{1440 F_{\pi }^4 s} \bigg[3 c_{\theta }^2 s_{\theta }^2 \left(-30 m_K^2m_{\eta '}^2-66 sm_{\eta '}^2+226 s
   m_K^2+15 m_K^4+15 m_{\eta '}^4\right)-16 \sqrt{2} s c_{\theta }^3 \left(12 m_K^2-7 m_{\pi }^2\right) s_{\theta }\nonumber\\&&+9 s\, c_{\theta }^2 s_{\theta }^2  \left(8 m_{\eta }^2-28 m_{\pi }^2+5 s\right)+8 \sqrt{2} s c_{\theta }
   \left(m_{\pi }^2-6 m_K^2\right) s_{\theta }^3+80 m_{\pi }^2 s c_{\theta }^4+80 s \left(m_{\pi }^2-3 m_K^2\right) s_{\theta }^4\bigg] \,.
\end{eqnarray}

\section*{Appendix B: the two-point one-loop function at finite temperatures}\label{sec.appendixB}

In this part, we discuss the evaluation of the two-point one-loop $G(s)$ function in Eq.~\eqref{eq.defg} at finite temperatures in detail. 

One can first separate out the integral of the zeroth component $k_0$ in Eq.~\eqref{eq.defg}
\begin{eqnarray}\label{eq.defgk0}
G(s)&=&-i\int\frac{{\rm d}^4k}{(2\pi)^4}
\frac{1}{(k^2-m_{1}^2+i\epsilon)[(P-k)^2-m_{2}^2 +i\epsilon]} \nonumber \\ 
&=&-i\int\frac{{\rm d}^3\vec{k}}{(2\pi)^3} \frac{d k_0}{2\pi}
\frac{1}{({k_0}^2-E_1^2+i\epsilon)[(P_0-k_0)^2-E_2^2 +i\epsilon ]}\,,
\end{eqnarray}
with
\begin{eqnarray}
s=P^2\,, \qquad P_\mu= (P_0, -\vec{P})\,,\qquad E_1^2= |\vec{k}|^2 + m_1^2\,, \qquad  E_2^2= |\vec{P}-\vec{k}|^2 + m_2^2\,.
\end{eqnarray}
In general cases Eq.~\eqref{eq.defgk0} depends on $P_0$ and $\vec{P}$ separately. In the CM frame of the two-body scattering, one has $\vec{P}=0$. 

We use the imaginary time formalism to include the finite-temperature contributions. This amounts to replacing the integration of the continuous $k_0$ with the discrete sum of $i\omega_n=i2\pi n T$~\cite{LeBellac}. In this way, one should take the substitution $k_0 \to i\omega_n$ and $dk_0 \to i 2\pi T$ in the last line of Eq.~\eqref{eq.defgk0}, which leads to 
\begin{eqnarray}\label{eq.gftsummid1} 
G(s)^{\rm FT}&=&-T\int\frac{{\rm d}^3\vec{k}}{(2\pi)^3} \sum_{n=-\infty}^{+\infty}
\frac{1}{(\omega_n^2+E_1^2)[(P_0-i\omega_n)^2-E_2^2 ]}\,, \nonumber \\
&=&-T\int\frac{{\rm d}^3\vec{k}}{(2\pi)^3} \sum_{n=-\infty}^{+\infty}
\frac{1}{2E_1}\bigg(\frac{1}{i\omega_n+E_1} - \frac{1}{i\omega_n-E_1} \bigg) \frac{1}{2E_2}\bigg(\frac{1}{i\omega_n-P_0-E_2} - \frac{1}{i\omega_n-P_0+E_2} \bigg)\,, \nonumber \\
&=& -T\int\frac{{\rm d}^3\vec{k}}{(2\pi)^3} \frac{1}{4E_1 E_2} \sum_{n=-\infty}^{+\infty}
\bigg[ -\frac{1}{P_0+ E_1 + E_2}\bigg( \frac{1}{i\omega_n+E_1} - \frac{1}{i\omega_n-P_0-E_2} \bigg) 
\nonumber \\  && 
+\frac{1}{P_0+ E_1 - E_2}\bigg( \frac{1}{i\omega_n+E_1} - \frac{1}{i\omega_n-P_0+E_2} \bigg)
+\frac{1}{P_0- E_1 + E_2}\bigg( \frac{1}{i\omega_n-E_1} - \frac{1}{i\omega_n-P_0-E_2} \bigg)
\nonumber \\  && 
-\frac{1}{P_0- E_1 - E_2}\bigg( \frac{1}{i\omega_n-E_1} - \frac{1}{i\omega_n-P_0+E_2} \bigg)
\bigg] \,,
\end{eqnarray}
where $P_0$ should take one of the possible $i\omega_n$ in the sum~\footnote{After explicitly performing the Matsubara sum, one can then analytically extrapolate the $P_0$ to other values~\cite{LeBellac}.}. 
In order to efficiently calculate the integral, it is necessary to evaluate the infinity sums by using the standard Matsubara techniques. The basic formula of the Matsubara sum is 
\begin{eqnarray}\label{eq.matsubarasum}
 T \sum_{n=-\infty}^{+\infty}  \frac{1}{i\omega_n \pm E} = \pm f(E) + \cdots\,,
\end{eqnarray}
where only the temperature-dependent terms are explicitly kept in the right side of the equation and the ellipses denote that the terms survive at zero temperature. By combining Eqs.~\eqref{eq.gftsummid1} and \eqref{eq.matsubarasum}, the temperature-dependent parts of the two-point one-loop function can be written as 
\begin{eqnarray}
 G(s)^{\rm FT}&=& \int\frac{{\rm d}^3\vec{k}}{(2\pi)^3} \frac{1}{4E_1 E_2}  
\bigg\{ \frac{1}{P_0+ E_1 + E_2+i\epsilon}\bigg[ f(E_1) + f(E_2) \bigg] +\frac{1}{P_0+ E_1 - E_2+i\epsilon}\bigg[ -f(E_1) + f(E_2)\bigg]
\nonumber \\  && 
+\frac{1}{P_0- E_1 + E_2+i\epsilon}\bigg[ f(E_1) - f(E_2) \bigg] -\frac{1}{P_0- E_1 - E_2+i\epsilon}\bigg[ f(E_1)+ f(E_2) \bigg]
\bigg\} \,,
\end{eqnarray}
which can be simplified to Eq.~\eqref{eq.gft} in the CM frame.

\end{document}